\documentclass[fleqn,usenatbib,useAMS]{mnras}

\usepackage{graphicx}	
\usepackage{amsmath}	
\usepackage{amssymb}	
\usepackage{multicol}        
\usepackage{bm}		
\usepackage{pdflscape}	

\usepackage[normalem]{ulem} 

\newcommand{\msun}{{M}_{\odot}}

\newcommand{\rhogc}{\rho_{\rm GC}}
\newcommand{\rhogcn}{\rho_{\rm GC0}}

\newcommand{\be}{\begin{equation}}
\newcommand{\ee}{\end{equation}}
\newcommand{\bi}{\begin{list}{\labelitemi}{\leftmargin=1em}\setlength{\itemsep}{-3pt}}
\newcommand{\ei}{\end{list}}

\newcommand{\beqn}{\begin{eqnarray}}
\newcommand{\eeqn}{\end{eqnarray}}

\newcommand{\rhn}{r_{\rm h,0}}

\newcommand{\pc}{{\rm pc}}

\newcommand{\gpc}{{\rm Gpc}}

\newcommand{\mlo}{m_{\rm lo}}
\newcommand{\mup}{m_{\rm up}}
\def\phicl{\phi_{\rm cl}}   
\def\phicln{\phi_{\rm cl,0}}   

\def\dr{{\rm d}}

\def\pc{{\rm pc}}
\def\gpc{{\rm Gpc}}

\def\msun{{\rm M}_\odot}
\def\Mc{M_{\rm c}}    

\def\Mc{M_{\rm c}}    
\def\mlo{m_{\rm lo}}    
\def\mup{m_{\rm up}}    
\def\Mlo{M_{\rm lo}}    

\def\rhn{r_{\rm h,0}}   

\usepackage[T1]{fontenc}
\usepackage{ae,aecompl}

\usepackage{txfonts,color}
\usepackage{soul}
\usepackage{subfig}

\newif\ifnew
\newtrue

\def\mgnew#1{}

\ifnew

\def\mgnew#1{{\it\textcolor{red}{\,MG:#1}}}
\fi
\definecolor{orange}{rgb}{1, 0.35, 0.}

\title[Black hole binary mergers in globular clusters]{
Coalescing black hole binaries from globular clusters: mass distributions and comparison to gravitational wave data from GWTC--3} 
\author[F. Antonini et al.]{Fabio Antonini\thanks{E-mail: AntoniniF@cardiff.ac.uk}$^{1}$,
Mark Gieles$^{2,3}$,
Fani Dosopoulou$^{4,5}$, and
Debatri Chattopadhyay$^{1}$
\\
$^{1}$Gravity Exploration Institute, School of Physics and Astronomy, Cardiff University, Cardiff, CF24 3AA, UK; \\ $^{2}$ {ICREA, Pg. Llu\'{i}s Companys 23, E08010 Barcelona, Spain}; \\ $^{3}${Institut de Ci\`{e}ncies del Cosmos (ICCUB), Universitat de Barcelona (IEEC-UB), Mart\'{i} Franqu\`{e}s 1, E08028 Barcelona, Spain} \\
$^{4}$ Princeton Center for Theoretical Science, Princeton University, Princeton, NJ 08544, USA\\
$^{5}$ Department of Astrophysical Sciences, Princeton University, Princeton, NJ 08544, USA
}

\begin{document}
\label{firstpage}
\pagerange{\pageref{firstpage}--\pageref{lastpage}}
\maketitle

\begin{abstract}
We use our cluster population model, {\tt cBHBd}, to explore the mass distribution of  merging black hole binaries formed dynamically in globular clusters. We include in our models the effect 
of mass growth through hierarchical mergers
and compare the resulting distributions
to those inferred from the third gravitational wave transient catalogue.
We find that none of our
models can reproduce the peak at $m_1\simeq 10M_\odot$ in the primary black hole mass 
distribution that is inferred from the data. This disfavours a scenario where most of the sources are formed in globular clusters. 
On the other hand, a globular cluster origin
can  account for the inferred secondary peak at $m_1\simeq 35M_\odot$, which  requires that the most massive clusters  form with half-mass densities
$\rho_{\rm h,0}\gtrsim 10^4~\msun\, \pc^{-3}\ $.
Finally, we find that the lack of a high mass cut--off  in the inferred mass distribution can  be  explained by the repopulation of an initial  mass gap through hierarchical mergers.  Matching the inferred merger rate above $\simeq 50M_\odot$ requires both initial cluster densities $\rho_{\rm h,0}\gtrsim 10^4~\msun\, \pc^{-3}\ $, and that  black holes form with nearly zero spin. A hierarchical merger scenario makes specific predictions for the appearance and position  of multiple  peaks in the black hole mass distribution, which can be tested against future data. 
\end{abstract}




\section{Introduction}
The   analysis of gravitational wave (GW) observations has identified  structures in the mass distribution of the observed population \citep{2021ApJ...913L..19T}. Some of these structures already emerged from the analysis of the second gravitational wave transient catalog \citep[GWTC-2;][]{2021ApJ...913L...7A,Abbott:2020niy}. However, thanks to the increased number of events 
in the new GWTC-3 \citep{2021arXiv211103634T,2021arXiv211103606T},
we are now more confident of their statistical significance.
In particular, three important features in the underlining  BH mass distribution have been uncovered:
(i) the   distribution of primary BH masses has a strong peak at about $\simeq 10M_\odot$; (ii) there is clear evidence for a secondary peak  at  $\simeq 35M_\odot$; and
(iii) there is no evidence  for any mass gap above $\approx 40$--$60M_\odot$, which is predicted by 
stellar evolution models due to pulsational pair--instability and pair--instability in massive stars \citep[e.g.,][]{Woosley2016,2017MNRAS.470.4739S,2022arXiv220409061O}. 
In this article, we perform a large number of cluster simulations to understand whether  (i), (ii) and (iii) can be explained by a globular cluster (GC) origin for the sources.

The formation of BH binary mergers, including those with components above the upper mass gap, might be explained by  several formation pathways. These include binary stellar evolution \citep[e.g.,][]{Dominik2012,Mink2015,Mandel2016a,2020ApJ...902L..36F,2021MNRAS.501.4514C}, multiple star interactions \citep[e.g.,][]{Silsbee2017,2017ApJ...841...77A,2020ApJ...895L..15F,2021MNRAS.502.2049L,2021MNRAS.506.5345H,2022arXiv220316544S}, stellar collisions in open clusters \citep[e.g.,][]{2020ApJ...903...45K,2020MNRAS.497.1043D,2021MNRAS.508.3045D,2021MNRAS.500.3002B,2022MNRAS.513.4527C},  primordial BHs \citep[e.g.,][]{2018JCAP...10..043B,2020JCAP...01..031G,2021PhRvL.126e1101D}, and formation in active galactic nuclei \citep[e.g.,][]{Bartos2016,Stone2016,2019PhRvL.123r1101Y,2021ApJ...908..194T}. One widely discussed scenario is formation through three body dynamical interactions  in dense stellar environments such as nuclear star clusters \citep[e.g.,][]{OLeary2009,2009ApJ...692..917M,2016ApJ...831..187A,2019MNRAS.486.5008A,2022arXiv220403745F} and GCs \citep[e.g., ][]{Sigurdsson1993,Kulkarni1993a,Banerjee2010,Rodriguez2015a,Rodriguez2016a,2017MNRAS.464L..36A,2018PhRvL.121p1103F,2022MNRAS.513.4527C}.

 The mass distribution of coalescing BH binaries produced in GCs has been investigated in several  studies \citep[e.g.,][]{Rodriguez2016a,2017MNRAS.464L..36A,2017MNRAS.469.4665P,2020PhRvD.102l3016A,2022MNRAS.511.5797M,2022arXiv220508549Z}.
Previous work suggests that GCs are an environment where BH binaries can efficiently assemble and merge, providing one of the main formation channels of BH binary coalescences in the Universe \citep{2000ApJ...528L..17P}.  
 In particular, it has been argued that due to the high escape velocities of GCs, BH mass growth 
can occur through consecutive mergers, populating any mass gap created by stellar processes \citep{2018PhRvL.120o1101R,2020ApJ...900..177K,2020ApJ...896L..10R,2020ApJ...893...35D,2021ApJ...915L..35K,2021MNRAS.507.3362T}.
In this scenario, a BH that is formed from a previous merger and is retained inside the cluster, sinks back to the cluster core where it dynamically couples with another BH and  merges with it after a series of binary-single encounters. If this process repeats multiple times, significant mass growth can occur \citep{,2016ApJ...831..187A,2017ApJ...840L..24F,2017PhRvD..95l4046G}.
A direct comparison of model predictions to data, however, {are rare} {\citep*{2022arXiv220303651M}}. It remains therefore an open question whether a GC origin provides a plausible explanation for the
inconclusive evidence for an upper mass gap in the GW data, and whether the other features of the inferred BH mass distribution can also be reproduced. {A putative successful GC model will then provide useful constraints on the properties of GCs and their BHs at birth.}

In this work, we adopt our new fast method {for the evolution of star clusters and their BBHs}, {\tt cBHBd} \citep{2020MNRAS.492.2936A}, to 
study the mass distribution of BHs  produced  dynamically in GCs, including the effect of hierarchical mergers {and a novel recipe for sampling masses of the BBH components and the  interlopers}. Our efficient approach allows us to address how model assumptions affect the final results, place error bars on merger rate estimates, compare to the distributions inferred from the new GW data catalog GWTC-3, and, finally, asses a hierarchical merger origin  for the formation of the most massive BHs detected by LIGO and Virgo. 

In Section \ref{model} we describe our methodology and approximations. In Section \ref{results} we describe our main results, and the importance of model assumptions.
We conclude and summarise our results in Section \ref{conc}.

\section{cluster models with hierarchical  mergers}\label{model}

We simulate the evolution of BH binaries in star clusters using our code {\tt cBHBd}, which we modify in order to include hierarchical mergers.
We define here hierarchical mergers as binary mergers in which at least one of the two BH components is a BH remnant that was formed from a previous merger.

 Our method is based on H\'{e}non's principle \citep{Henon1975}
which states that the rate
of heat generation in the core is a constant fraction of the
total cluster energy per half-mass relaxation time. Thus, the energy production rate in the core, which we assume is produced by BH binaries, is regulated by the energy demand of the entire system \citep{2013MNRAS.432.2779B}. The lifetime and the merger rate of BHs in the cluster can be linked to the evolution of the cluster itself as described in details in \citet{2020MNRAS.492.2936A}. Then, three  ingredients are needed in order to determine the 
 formation of BH binaries, their merger rate and their properties: (i) a model for the evolution of the cluster global properties; (ii) a model for binary BH dynamics; and (iii) a realistic set of initial BH masses.
 
 We start by sampling the masses of the stellar progenitors of BHs from a standard mass function, $\phi(m_\star)\propto m_\star^{-2.3}$ \citep{1955ApJ...121..161S,Kroupa2001a}, with masses in the range $20M_\odot$ to $130M_\odot$.
 For 
 a given cluster metallicity, $Z$,
we evolve the stars to BHs using the Single Stellar Evolution
($\texttt{SSE}$) package \citep{2002MNRAS.329..897H}, which we modified to include  updated prescriptions for stellar winds and mass loss \citep[following][]{Vink2001}, and for pair--instability in massive stars \citep[following][]{2017MNRAS.470.4739S}. We therefore evolve the BH progenitors as single stars, assuming a zero binary fraction initially.
{   At the end of this phase, the total number of BHs in a cluster model is calculate by assuming a \citet{,Kroupa2001a} initial mass function in the mass range  $0.1M_\odot$ to $130M_\odot$.}
 The value of the largest BH mass formed in the model depends on metallicity and varies between 
 $\simeq 25M_\odot$ for $Z=2\times10^{-2}$ and $\simeq 55M_\odot$  for
 $Z=1\times10^{-4}$.
For each BH we compute a natal kick velocity from a
Maxwellian distribution with dispersion $265{\rm km\ s^{-1}}$ \citep{Hobbs2005}, lowered by the amount of  mass that falls back into the forming compact object \citep{Fryer2012}.
In most of our models, we start the BHs all with the same value of the spin angular momentum, $\chi$, where $\vec{\chi}=\vec{S}/m^2$ is the
dimensionless spin of the BH  and $\vec{S}$ is the spin angular momentum in units of $m^2$. In one model the initial value of $\chi$ is sampled from a distribution that is consistent with that inferred from the GW data and is given by the median distribution shown in figure~15 of \cite{2021arXiv211103634T}.

Then, we initialise and evolve the cluster model. The initial conditions are determined by three parameters:
the  cluster density, $\rho_{\rm h,0}$; the  cluster mass, $M_{\rm 0}$; and the  total mass in BHs,   $M_{\rm BH,0}$.
The latter is set equal to the total mass in
BHs obtained with $\texttt{SSE}$, assuming a Kroupa initial mass function in the range $0.08-130\ M_\odot$  and taking into account that a fraction of the BHs are ejected from the cluster by a natal kick.
The time evolution of the cluster properties is then obtained as in \citet{2020PhRvD.102l3016A}. Briefly, we integrate a set of first order differential equations which determine the time evolution of $M_{\rm}$,  $M_{\rm BH}$, and the cluster half mass radius, $r_{\rm h}$. These models include simple prescriptions for mass loss {and expansion} due to stellar evolution and cluster `evaporation', while BHs are assumed to be lost through dynamical ejections.

 Finally, we dynamically evolve the  BH binaries that form via three-body processes in the cluster core. Our treatment of binary BH formation and evolution follows closely  \citet{2020MNRAS.492.2936A}.  The first binary BH forms after the cluster core-collapse time 
\begin{equation}\label{cc}
\tau_{\rm cc}= 3.21 t_{\rm rh,0}
\end{equation}
where $t_{\rm rh,0}$ is the initial cluster half-mass relaxation time {\citep[for the definition, see equation~10 in][]{2020MNRAS.492.2936A}}.
We assume that the binary is formed with a semi-major axis at the soft-hard boundary,
$a_{\rm h}\simeq {G \mu /\sigma^2}$, with $\mu=m_1m_2/(m_1+m_2)$,
where $m_1$ and $m_2$ are the masses of the binary components, and $m_1>m_2$.
The expression of $a_{\rm h}$ above is only approximate, and valid under the assumption of equal mass components.  Later  in Appendix\ \ref{masses} we introduce the quantity $\beta$ and equipartition among BHs of different masses, then the definition is $a_{\rm h}=0.5Gm_1m_2\beta$. 

{   The pairing of BHs is done by sampling  their masses from the set of BHs still left inside the cluster.
We first draw two mass values from  the power law probability distributions 
$p(m_1)\propto m_1^{\alpha_1}$ and $p(q)=q^{\alpha_2}$, with $\alpha_1=8+2\alpha$,  $\alpha_2=3.5+\alpha$ and $q=m_2/m_1$. 
 Here $\alpha$ is the power law index of the  BH mass function, which also evolves with time as the BH population is depleted.
 The two BH components are then selected by 
choosing the two BHs that have the mass closest to the values drawn from $p(m_1)$ and $p(q)$ (or $p(m_2)$).
 
Once selected, the binary is evolved through a sequence of binary-single encounters.
Similarly, we find the mass of the third BH interloper from  the power law distribution $p(m_3)\propto m_3^{\alpha_3}$ with $\alpha_3=\alpha+1/2$. 
The adopted expressions for $p(m_1)$, $p(q)$ and $p(m_3)$ are motivated below in Appendix\ \ref{masses}.
The power law exponent, $\alpha$, is obtained at the start of the integration for each cluster from a fit to the initial BH mass function after removing BHs that are ejected by natal kicks.
The value of $\alpha$ as well as the lower and upper bound of the BH mass function are then recalculated after each time-step.
 Specifically, the lower bound of the BH mass function is set equal to the mass of the lightest BH in the cluster, and the upper bound is the mass of the most massive BH. This procedure allows to take into account
the evolution of the BH mass function with time due to ejections and the growth of BHs through hierarchical mergers.}

 Following \citet{2017arXiv171107452S}, we divide each binary-single encounter in a set of $N_{\rm rs}=20$ resonant intermediate states and assume  that the eccentricity of the binary after each state is sampled from a thermal distribution $N(<e)\propto e^2$. 
If 
\begin{equation}\
\sqrt{1-e^2}< h\left(R_{\rm S}\over a \right)^{5/14}
\end{equation}
a merger occurs through a GW capture before the next intermediate binary-single state is formed,
where $R_{\rm S}=2G(m_1+m_2)/c^2$ and $h$ is a constant of order unity.  

If the binary survives  the 20 intermediate resonant states, we
compute:
 (i) the new binary semi-major axis, assuming that
 its binding energy decreases by the fixed fraction
 $\Delta E/E=0.2$ \citep{2017arXiv171107452S}.
 (ii) the recoil kick due to energy and angular momentum conservation experienced by the binary centre of mass \citep{2016ApJ...831..187A}
\begin{equation}
v_{\rm bin}^2=0.2G{m_1m_2\over m_1+m_2+m_3}q_3/a
\end{equation}
with $q_3=m_3/(m_1+m_2)$,
and (iii) the recoil kick experienced by the interloper:
\begin{equation}
v_{\rm 3}=v_{\rm bin}/q_3\ .
\end{equation}
If $v_{\rm bin}>v_{\rm esc}$, the binary is ejected from the cluster; if $v_{\rm 3}>v_{\rm esc}$, the interloper is also ejected from the cluster. If the binary is ejected from the cluster, we compute its merger timescale due to GW energy loss using the standard Peter's formula \citep{Peters1964}.

If  $v_{\rm bin}<v_{\rm esc}$, 
and the binary angular momentum 
 at the end of the triple interaction is such that \citep{2020MNRAS.492.2936A}
\begin{equation}\label{merger}
\sqrt{1-e^2}< 1.3 \left[G^4 (m_1m_2)^2(m_1+m_2)\over c^5 \dot{E}_{\rm bin} \right]^{1/7} a^{-5/7}\ ,
\end{equation}
then the BH binary merges before the next binary-single encounter takes place. Binaries that undergo this type of evolution are often named  `in-cluster inspirals' \citep{2017arXiv171107452S,2018PhRvL.120o1101R}. We then assign the new remnant BH a GW recoil kick, $v_{\rm GW}$, 
 and compute its new spin and mass following \cite{2008PhRvD..78d4002R}. 
If $v_{\rm GW}>v_{\rm esc}$ the remnant is ejected from the cluster, otherwise we compute the dynamical friction timescale to sink back to the cluster core
\begin{equation}\label{dfr}
\tau_{\rm df} \simeq 1.65\,r_{\rm in}^2{\sigma\over {\ln \Lambda G m}}
\end{equation}
where \citep{2019MNRAS.486.5008A}
\begin{equation}\label{dfr2}
r_{\rm in}=r_{\rm h}\sqrt{{v_{\rm esc}^4\over 
\left(v_{\rm esc}^2-v_{\rm GW}^2\right)^2
}
-1}
\end{equation}
 and only allow the BH to form a new binary after this time.
If $v_{\rm bin}<v_{\rm esc}$,  but condition equation~(\ref{merger}) is not satisfied, then a new interloper is sampled from the given distribution and the binary is evolved through a new binary-single interaction. 

Each binary is evolved through a sequence of binary-single encounters until either a merger occurs or it is ejected from the cluster. 
Then  a new binary is formed.
The cluster is assumed to live in a state of balanced evolution in which  the binary disruption rate is equal to the binary formation rate. Under this assumption, the lifetime of a binary, or the timescale until a new binary is formed, is simply
\begin{equation}
\tau_{\rm bin} = {m_{\rm ej}\over \dot{M}_{\rm BH}}, 
\end{equation}
where $m_{\rm ej}$ is the total mass ejected by the binary, and $\dot{M}_{\rm BH}$ is the BH mass loss rate given by the cluster model.

We continue selecting new binaries and evolve them through binary-single encounters until either all BHs have been ejected from the cluster, or until a maximum integration time of $t=13\rm Gyr$ has passed.
  
\begin{figure*}
\includegraphics[width=0.35\textwidth]{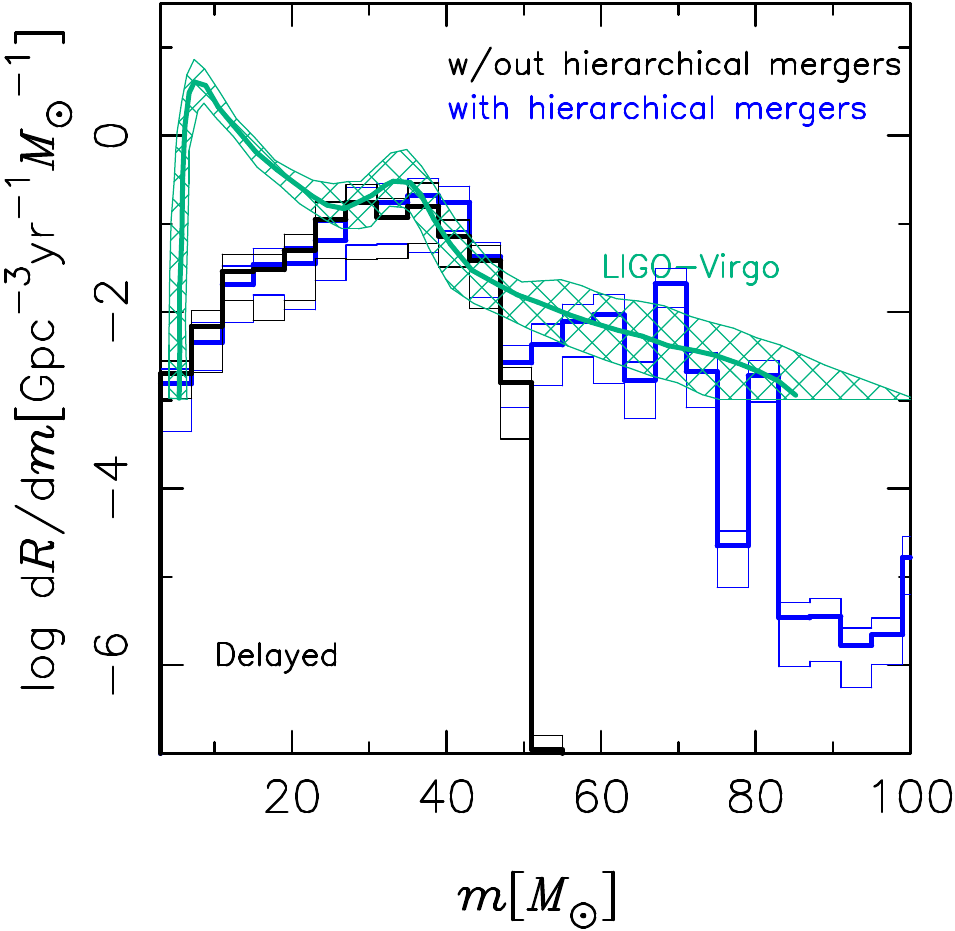}
\includegraphics[width=0.34\textwidth]{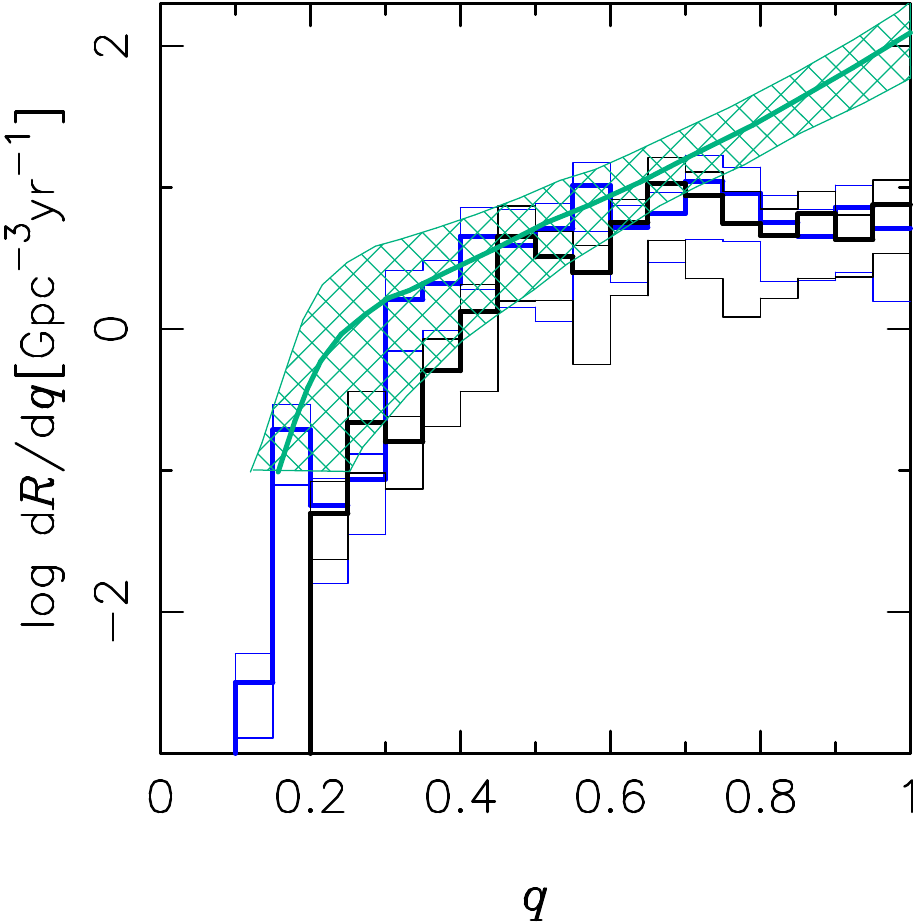}
\includegraphics[width=0.35\textwidth]{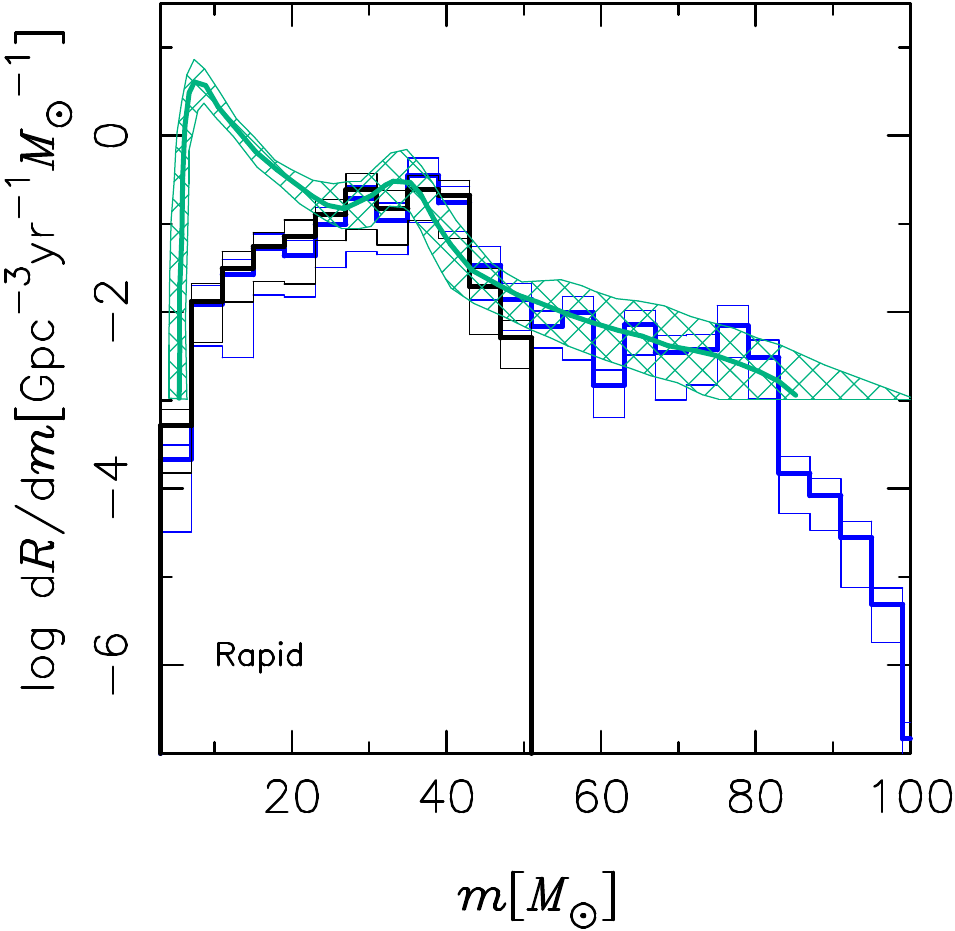}
\includegraphics[width=0.34\textwidth]{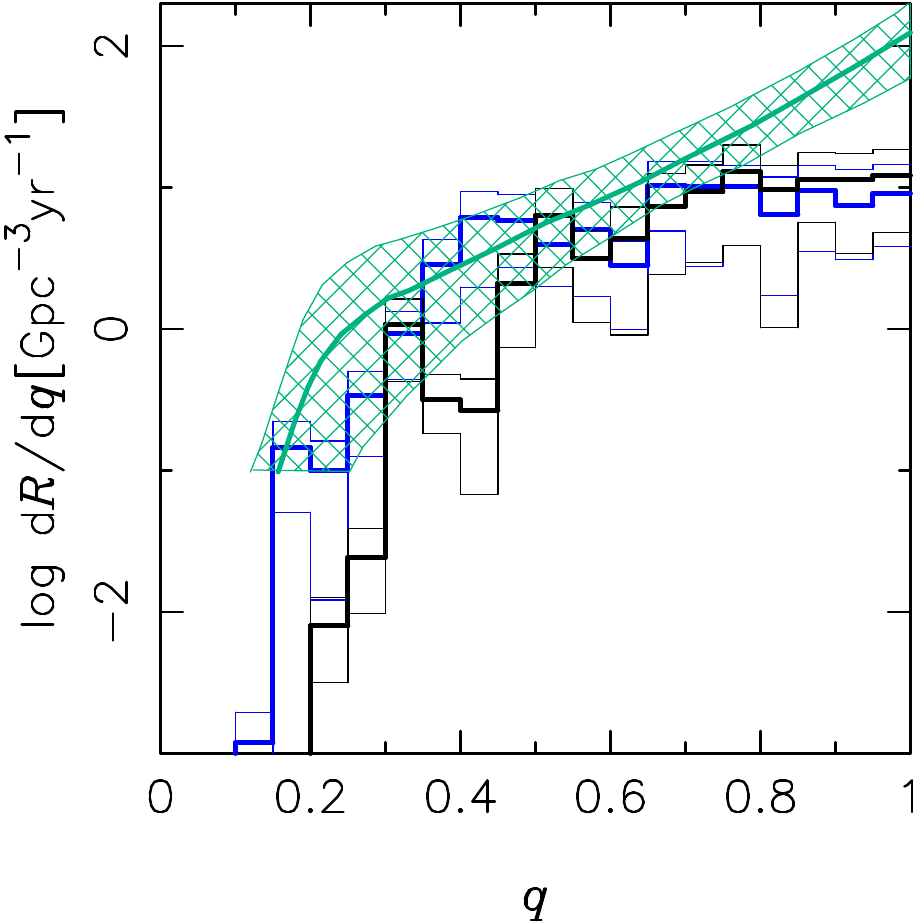}
\includegraphics[width=0.35\textwidth]{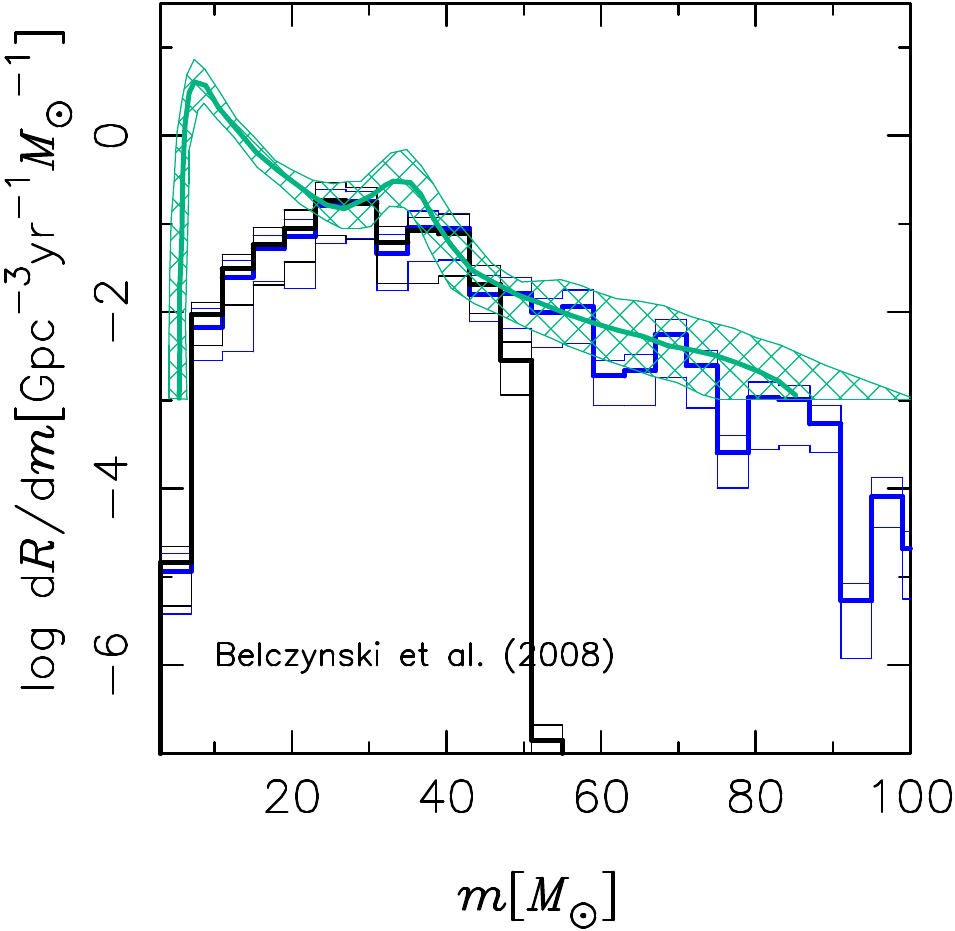}
\includegraphics[width=0.34\textwidth]{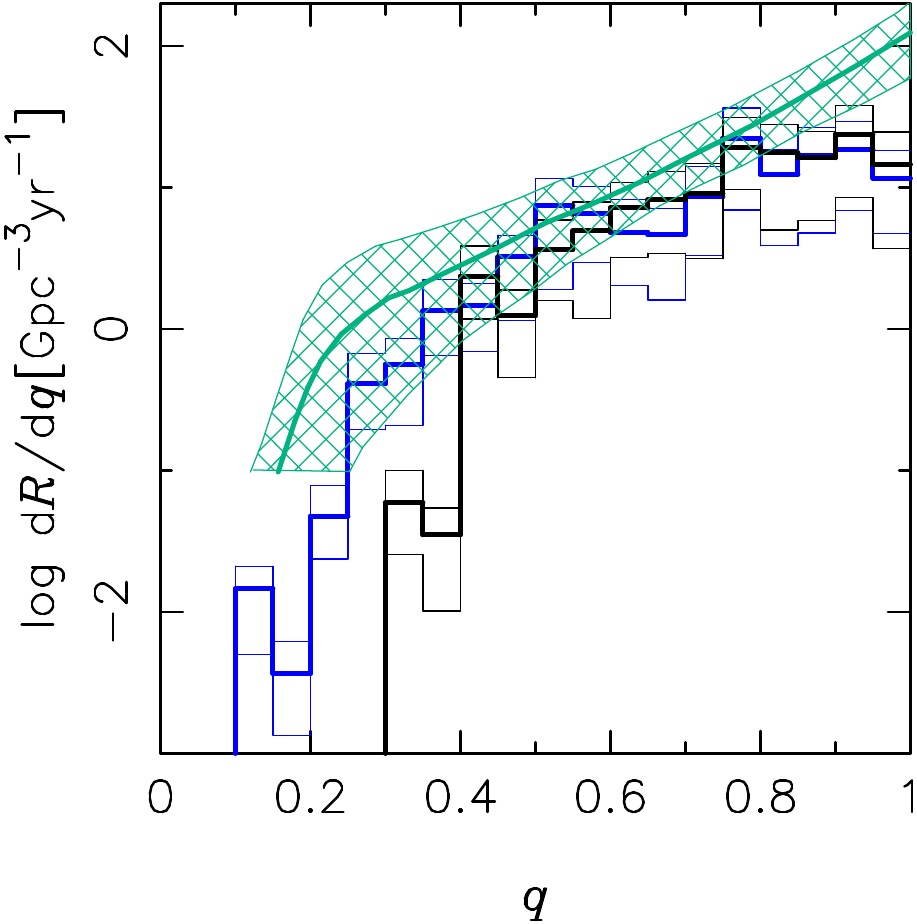}
\caption{Local distributions of primary BH mass and mass ratio for merging BH binaries in our best-case scenario where clusters start with high densities,
$\rho_{\rm h ,0}=10^5~\msun\, \pc^{-3}\ $, and  the BHs are all initialised with zero dimensionless spin parameter $\chi=0$.
Top, middle and bottom panels correspond to the delayed supernova mechanism, the rapid supernova mechanism, and the BH mass distribution of \citet{Belczynski2008}, respectively.
Black lines show the corresponding distributions when hierarchical mergers are not included in the calculation, i.e., it is assumed that a  BH formed from a previous merger is always ejected from the cluster. Thick lines show median values of the merger rate value, and thin lines the corresponding $90\% $ confidence intervals. The green lines and hatched regions show the median and corresponding $90\% $ confidence regions inferred from the GWTC-3\citep{2021arXiv211103606T}.
}\label{nospin}
\end{figure*}

\subsection{Cluster initial mass function, formation time, and metallicity}
In order to generate predictions for  BH binary mergers, we need a GC initial mass function, and a model for how the formation rate and metallicity of clusters evolve with redshift.  These  ingredients of our models are described below. {For this we follow the approach of \citet{2020PhRvD.102l3016A}.}

The cluster initial mass function is obtained by fitting an evolved Schechter mass function to the observed
GC mass function in the Milky Way today
\citep{2007ApJS..171..101J}
\begin{equation}
\label{eq:evsch}
\phicl = A(M+\Delta)^{-2}\exp\left(-\frac{M+\Delta}{\Mc}\right) \ .
\end{equation}
This gives the posterior distribution for the parameters $\Mc$ and $\Delta$.
Adopting a simple model
for cluster evaporation and mass loss due to stellar evolution,
the corresponding initial GC mass function is given by:
\begin{equation}
  \phicln = 2A M_0^{-2}\exp\left(-\frac{M_0}{2\Mc}\right) .
\label{CIMF}
\end{equation}
The corresponding fractional mass loss due to evaporation and stellar evolution is 
\begin{equation}\label{fml}
K=  \frac{\rhogcn}{\rhogc} = \frac{\int_{\Mlo}^{\infty}\phicln M_0\dr M_0}{\int_{\Mlo}^{\infty} \phicl M\dr M} \ =32.5^{+86.9}_{-17.7} \ \rm (90\% \ cred.\ intervals) .
\end{equation}
The spread in $K$ provides an estimate of the uncertainty in the fractional mass loss from cluster until today, given the  156  Milky Way GC masses.
The cluster mass formed per unit volume integrated over all times is \citep{2020PhRvD.102l3016A}
\begin{align}
  \rhogcn 
   =  2.4^{+2.3}_{-1.2}\times10^{16}~\msun\,\gpc^{-3}\ \rm (90\% \ cred.\ intervals) .
   \label{eq:rhogcn}
\end{align}
The large error bars here  imply that $\rhogcn$ is uncertain by a factor of $\simeq 2$. In the next sections we include this uncertainty as well as the uncertainty on $K$ in the predictions for the merger rate.

We  obtain the distribution of cluster formation times
from the semi-analytical galaxy formation model of \citet{2019MNRAS.482.4528E}.
 The resulting cluster formation history peaks at a redshift of $\sim4$, which is earlier than the peak in the cosmic star formation history (redshift $\sim2$, \citealt{Madau2014}).  We sample the formation redshift of our cluster models from the total cosmic cluster formation rate given by 
the fiducial model of \citet{2019MNRAS.482.4528E}
  and integrated over all halo masses. This corresponds to  the formation rate per comoving volume of their Fig.~8 with
their parameters $\beta_\Gamma=1$ and $\beta_\eta=1/3$, where $\beta_\Gamma$ sets the dependence of the cluster formation efficiency on surface density, and $\beta_\eta$ the dependence
of the star formation rate on the halo virial mass.
Here, we renormalise the cluster formation rate, $\phi_{z}(z)$, such that $\int_\infty^0 \phi_{z}\dr z = 1$. Thus,
 we only sample the cluster formation time from \citet{2019MNRAS.482.4528E},  and then rescale  the cluster formation rate such to reproduce the total mass density given by our equation\ (\ref{eq:rhogcn}).
We note that the cluster formation model has a negligible impact on the local merger rate and properties of the merging binaries. In \citet{2020PhRvD.102l3016A} we showed that unrealistic models where all clusters are assumed to form at the same time (e.g., $z=3$) produce similar results than models in which the cluster formation rate is varied with redshift.

For the cluster metallicity,  we fit a quadratic polynomial to the observed age-metallicity relation for
the Milky Way GCs \citep{VandenBerg2013}, to obtain the mean metallicity
\begin{equation} \label{mmet}
\log(Z_{\rm mean}/Z_\odot)\simeq0.42+0.046\left({t\over {\rm Gyr}}\right)-0.017\left({t\over {\rm Gyr}}\right)^2 \ .
\end{equation}
Given the cluster formation redshift, we then assume a log-normal distribution of metallicity around the mean
\begin{equation} \label{metdis}
    \phi_Z=\frac{\log(e)}{\sqrt{2\pi\sigma^2}}\exp\Big\{-\frac{[\log(Z/{\rm Z_\odot})-Z_{\rm mean}]^2}{2\sigma^2}\Big\},
\end{equation}
with standard deviation $\sigma= 0.25$\,dex. This
takes into account the large spread found in the observed age-metallicity relation for the Milky Way GCs.

\subsection{Merger rates and their error bars}
Finally, we  construct a library of cluster models over a 3-dimensional grid of formation time, metallicity and cluster mass.

We sample the cluster formation redshift over the range $z\in [10;0]$ with 
step-size $\Delta z=0.5$; at a given redshift, the metallicity of the cluster is 
sampled in the range $Z\in [10^{-4};0.02]$ with logarithmic step size $\Delta \log Z= 0.1$; finally, for a given formation time and metallicity, the initial mass of the cluster is varied in the range $M_{\rm 0}\in [10^{2};2\times 10^7]M_\odot$, with step size $\Delta \log M_0/M_\odot= 0.1$.
The merger rate is then calculated over the grid of cluster models as:
\begin{equation}\label{mrate}
\mathcal{R}(\tau)=K\rhogc {\sum\limits_{z}^{} \sum\limits_{Z}^{}  \sum\limits_{M_0}^{}\dot{\mathcal{N}}(\tau; z, Z, M_{0})\phi_{z} \phi_{Z} \phicln M_{0}
\over
\sum\limits_{z}^{}\phi_{z}
\sum\limits_{Z}^{}\phi_{Z}
\sum\limits_{M_0}^{}\phicln M_{\rm 0}^2
},
\end{equation}
where $\dot{\mathcal{N}}(\tau; M_0, \rhn, Z)$ is the BH binary merger rate at a look-back time $\tau$ corresponding to a cluster
with an initial mass $M_0$,  metallicity $Z$ and
that formed at a redshift $z$.

In order to take into account the uncertainties in the initial cluster mass function, we sample 100 values over the posterior distributions of the parameters $M_{\rm c}$ and $\Delta$ obtained from the MCMC fit to the Milky Way GC mass function. 
 We also take into account the uncertainty on the  mass density of GCs in the Universe, $\rhogc$.
We assume that the parameter
$\rhogc$ follows a Gaussian distribution with mean $7.3\times 10^{14}M_\odot\ \rm Gpc^{-3}$ and dispersion $\sigma=2.6\times 10^{14}M_\odot\ \rm Gpc^{-3}$. We sample 100 values from this Gaussian distribution and for each of them we use
 equation~(\ref{mrate}) to determine a merger rate estimate for each of the [$\Mc$, $\Delta$] values, 
and thus obtain a {\it distribution} of merger rate density values.
 Since in this work we are interested in the mass distribution of local BH binary mergers, we   consider the differential merger rate in the local universe $\dr \mathcal{R}(z=0) / \dr m_1$ and $\dr \mathcal{R}(z=0) / \dr q$, which we compare to the distributions inferred from GWTC-3.

\begin{figure}\centering
\includegraphics[width=0.43\textwidth]{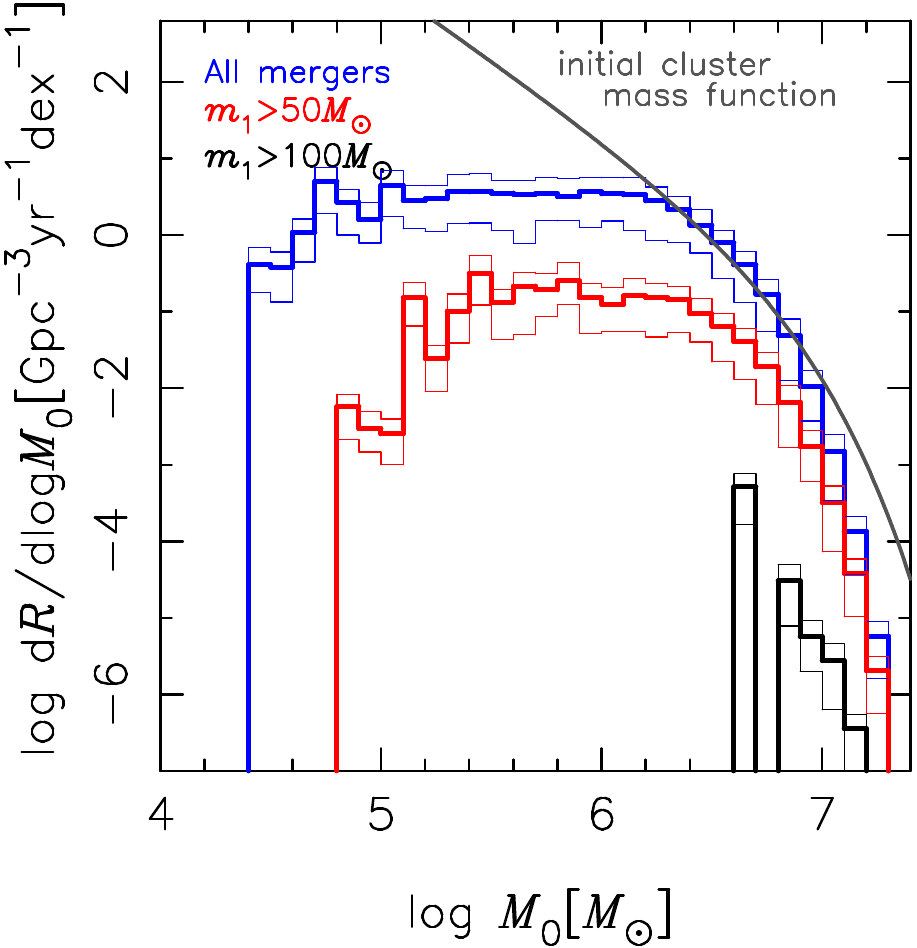}
\caption{Differential local merger rate as a function of the 
 initial cluster mass. We also show the initial cluster mass function (in arbitrary units) for our best fit value of Schechter mass, $\log M_{\rm c}/M_\odot=6.26$. The delayed supernova mechanism has been adopted here.
 }\label{clusterD}
\end{figure}

\begin{figure*}
\includegraphics[width=0.35\textwidth]{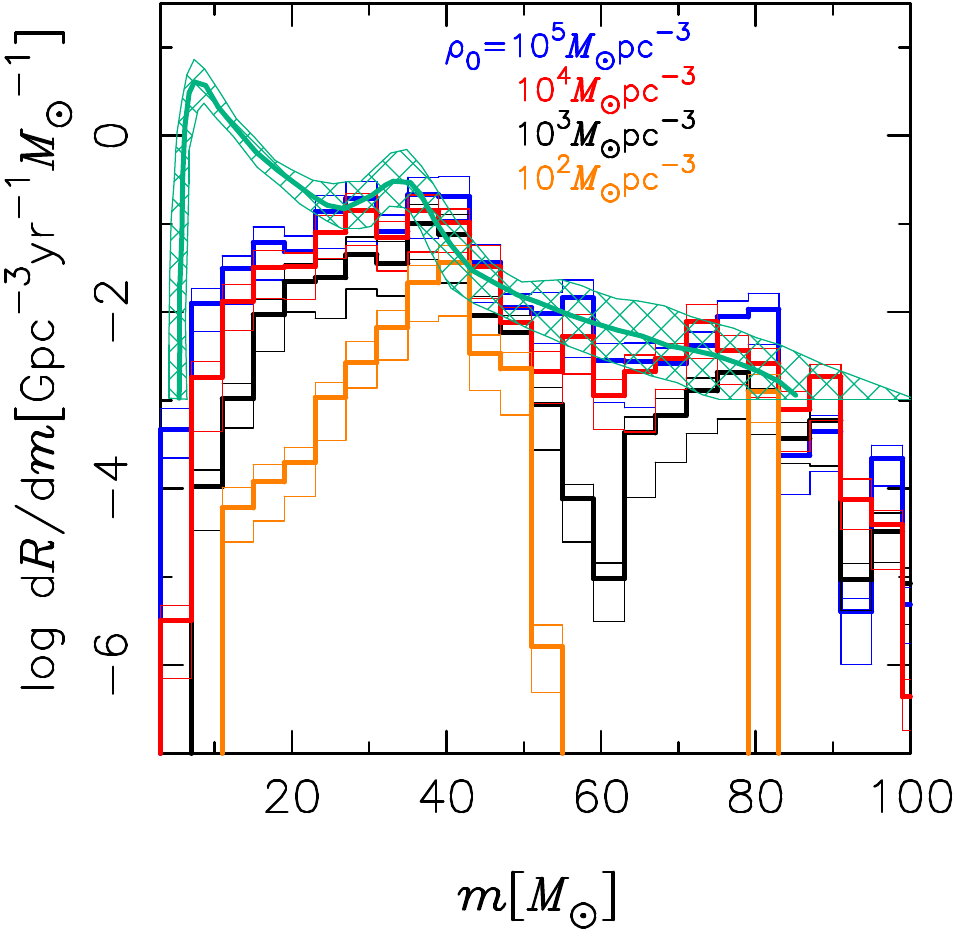}
\includegraphics[width=0.34\textwidth]{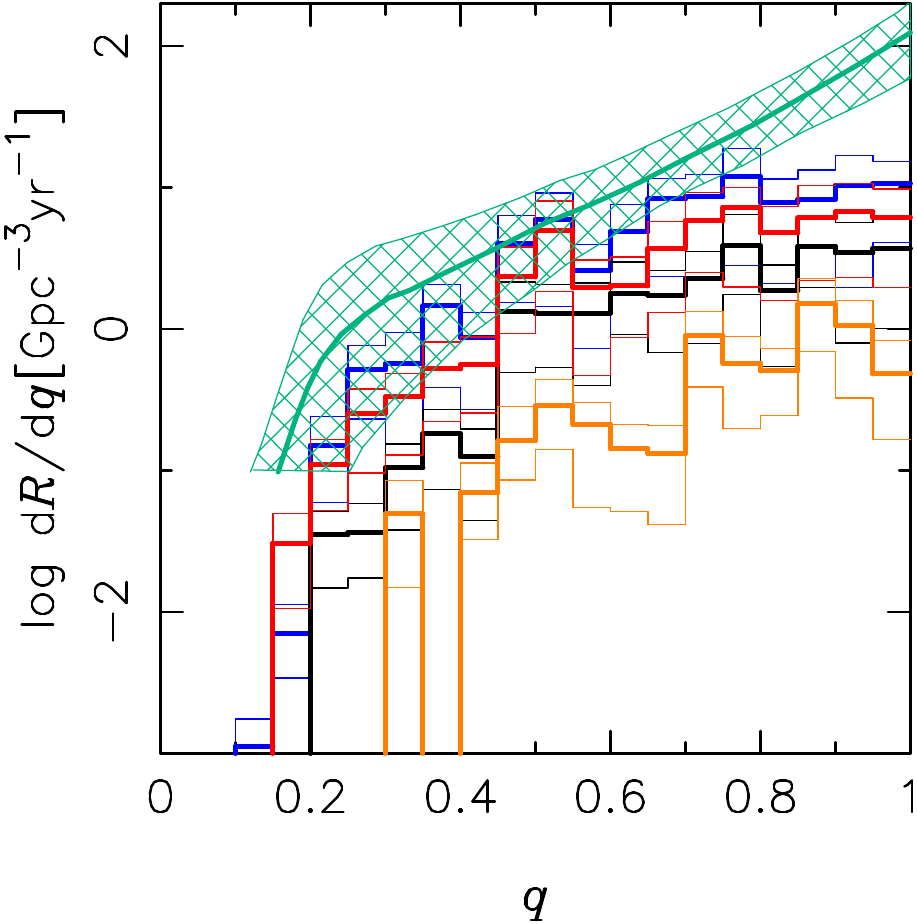}
\includegraphics[width=0.35\textwidth]{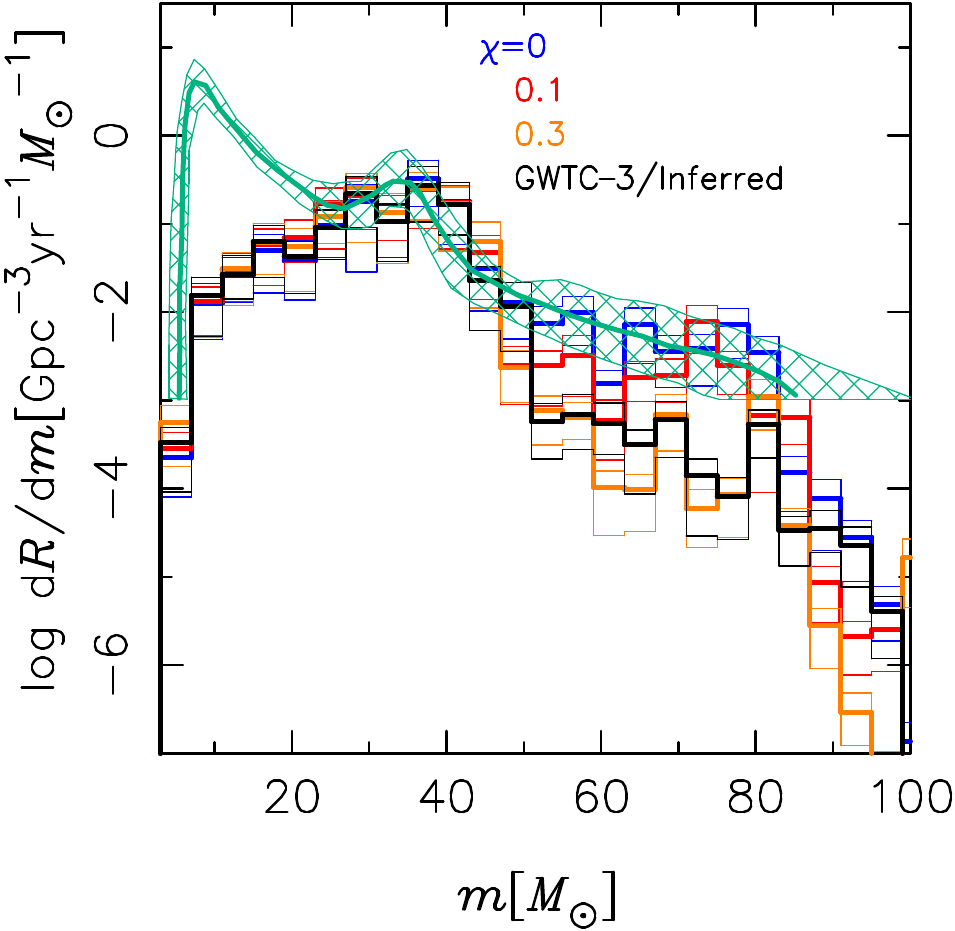}
\includegraphics[width=0.34\textwidth]{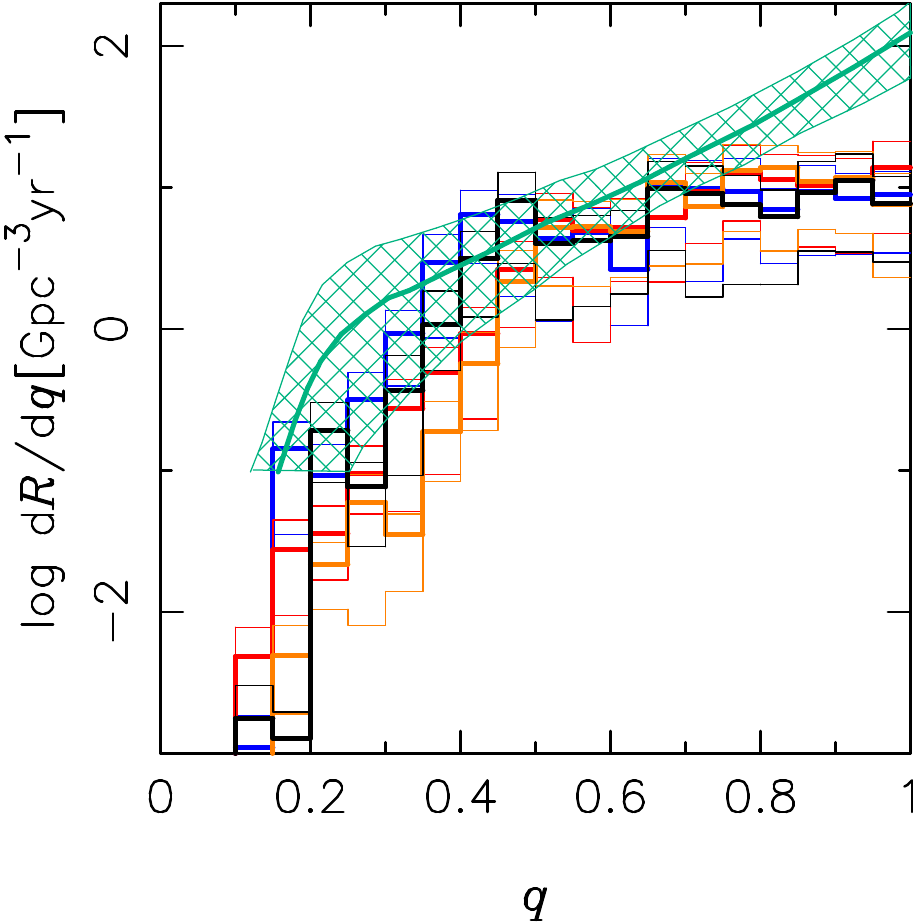}
\caption{Dependence of mass and mass ratio distributions 
on initial cluster half-mass density, and initial BH spins. The delayed supernova model is assumed here. Top panels use $\chi=0$
and the half-mass density of the cluster is varied as indicated. In the bottom panel we take $\rho_{\rm h ,0}=10^5~\msun\, \pc^{-3}$ and change the initial value of $\chi$; the black histograms show the results for a model where the initial value of $\chi$ is sampled using the inferred distribution of BH spins shown in Fig.~15 of \citet{2021arXiv211103634T}. In all the other models, the BHs all form with the same value of $\chi$ as indicated.
}\label{density}
\end{figure*}

\section{Results}\label{results}

\subsection{Primary BH mass and mass ratio distributions}\label{fidmod}
In \cite{2020PhRvD.102l3016A} we  found good agreement between model predictions and the inferred  distribution of primary BH masses within the range of values $m_1=15M_\odot$ to $40M_\odot$. Outside this range, the binary BH merger rate was found to be several orders of magnitude smaller than inferred. The first question we address here is whether the inclusion of hierarchical mergers can reduce the discrepancy at  $m_1\gtrsim 40M_\odot$ between models and the inferred astrophysical distributions. 

In Fig.~\ref{nospin} we plot the  distributions of $m_1$ and $q$ for three different assumptions about the initial BH mass function. In the upper panels we use the delayed supernova mechanism, in the middle panels the rapid supernova mechanism \citep{Fryer2012}, and in the lower panels we use the BH mass distribution from
\cite{Belczynski2008}.                  
These prescriptions produce somewhat different initial BH mass functions, and lead also to different natal kick values. 
In these models, all clusters are initialised with the same half-mass radius density of
$\rho_{\rm h ,0}=10^5~\msun\, \pc^{-3}\ $ and  the BHs are all started with zero dimensionless spin parameter, $\chi=0.$

In the left panels of Fig.~\ref{nospin}, we see that the new models produce  mergers above the $\sim 50M_\odot$ threshold.
These mergers are produced by BHs that grow hierarchically through mergers -- the vast majority of them are mergers between a first generation BH and a second generation BH. When we include these mergers in our calculation, we find good agreement between the models and the inferred distributions at $m_1\gtrsim 50M_\odot$, in the sense that the models give a merger rate that is comparable to the inferred value.
However, we also note that a simple power-law profile above this mass threshold is not a good representation of the model distributions.  Above $m_1\simeq 50M_\odot$, the model distributions are characterised by several peaks. Such higher mass peaks are related to peaks in the  BH mass distribution at lower masses. From Fig.~\ref{nospin} we see that the merger rate between first generation BHs peaks at $\simeq 30M_\odot$ and $40M_\odot$. Thus, mergers between first and second generation BHs lead to additional peaks at $\simeq (30+30)M_\odot=60 M_\odot$,  $(30+40)M_\odot=70 M_\odot$ and
$(40+40)M_\odot =80M_\odot$. The presence of peaks within the pair--instability mass gap
and their relation to  lower mass peaks in the BH mass distribution
is a clear prediction of a hierarchical  merger model for the origin of the binaries.

The black histograms in the left panels of Fig.~\ref{nospin}
show the results from models in which any remnant BH formed from a previous merger is   ejected  from the cluster `by hand'.
 In these models,  the distributions are sharply truncated at $\sim 50M_\odot$ since BHs cannot grow hierarchically above this mass value.
The  merger rate at $m_1\simeq 10M_\odot$ derived from all models is about two orders of magnitude smaller than the inferred rate.
This lower-mass peak can be explained, however, through other scenarios, including formation in the galactic field 
\citep[e.g.,][]{2022MNRAS.tmp.1776B,2022arXiv220913609V} and formation in young and open star clusters {because of their higher metallicity} \citep[e.g.,][]{2021MNRAS.503.3371B,2022MNRAS.513.4527C}.
On the other hand, our models reproduce the inferred merger rate near  $m_1\simeq 30M_\odot$, which can therefore be explained by a GC origin. This peak in the  mass distribution is due to mergers involving first-generation BHs, and it is  not related to hierarchical mergers.
 
 
{   By comparing the results in Fig.~\ref{nospin} with the  models in \cite{2020PhRvD.102l3016A}, we find that the latter generated a   merger rate  at $m_1\lesssim 20~M_\odot$ higher by a factor $\lesssim 2$}. The reason for this difference is due to the adopted new recipe for sampling the  black hole binary components and the interloper masses. In \cite{2020PhRvD.102l3016A} we had assumed that $m_1=m_2=m_3${=$m_{\rm max}$, where $m_{\rm max}$ is the mass of the most massive BH in the cluster}. The  distributions in Appendix~\ref{masses} mean instead that in the current models  $\langle m_3 \rangle\ll \langle m_1 \rangle\simeq  \langle m_2 \rangle$. Thus, each binary ejects more  low-mass BH interloopers lowering the overall BH merger rate at low masses.
 
In the right panels of  Fig.~\ref{nospin} we consider the distribution of the mass-ratio $q$. The new models result in a significantly higher rate of merging binaries  with small mass ratio, $q\lesssim 0.5$, providing a better match to the inferred distribution than models without hierarchical mergers. 
Most of these additional low-$q$ systems are  mergers between a first generation BH and a BH that formed through a previous merger  above the pair-instability mass limit.
 At high values of $q$, instead, both models with and without hierarchical mergers
produce a similar merger rate, which, at $q\gtrsim 0.8$, is  about one-order of magnitude smaller than inferred from the data. 

In Fig. \ref{clusterD} we  show the differential contribution to the local merger rate with respect to the initial cluster mass. This allows us to identify 
in which  type of clusters most of the mergers are formed.
The contribution to the total merger rate is nearly constant {   in log bins} between $M_0=5\times 10^4M_\odot$ and $5\times 10^6M_\odot$, while it  decrease exponentially above this range because of the 
truncation of the initial GC mass function at $\log M_{\rm c}/M_\odot \simeq 6.26$.
We then show the same cluster mass distribution, but only considering  mergers with $m_1\ge 50M_\odot$ (red histogram). These mergers, involving a primary BH above the mass gap limit, are mostly produced in  clusters with relatively large masses, between  $\sim 2\times 10^5M_\odot$ and $5\times 10^6M_\odot$. The fraction of these higher mass mergers to the total number of mergers  in each cluster mass bin increases with cluster mass.  By comparing the blue and the red histograms in the figure, we see that at $M_0\sim 10^7M_\odot$, between $ 10$ to $30\%$ of mergers have a primary with $m_1>50M_\odot$. The percentage goes down to $\sim 1\%$ in   clusters with an initial mass lower than $\simeq 10^6M_\odot$. Finally, we  show the $M_0$ distributions for the most massive mergers produced in our models, $m_1\ge 100M_\odot$.  These  BHs 
originate from at least two previous mergers since their mass is larger than twice the initial mass cut-off at $\simeq 50M_\odot$.
These systems  are formed in the most massive GCs, with initial mass well above the Schechter mass.  
  
Based on  the models of Fig. \ref{clusterD}, we compute a local binary BH merger rate of
$4.1^{+2.2}_{-2.5}\rm Gpc^{-3}~yr^{-1}$ (delayed), $4.5^{+2.7}_{-2.9} \rm Gpc^{-3}~yr^{-1}$ (rapid), and $6.0^{+3.6}_{-4.0}\rm Gpc^{-3}~yr^{-1}$ \citep[BH mass distribution from][]{Belczynski2008}, at $90\%$ confidence. The binary BH merger rate inferred from the gravitational wave data is estimated to be between $17.9 \rm Gpc^{-3}~yr^{-1}$ and 
$44 \rm Gpc^{-3}~yr^{-1}$ \citep{2021arXiv211103606T}, and it is therefore a factor of $\simeq 2$ to $20$ larger than the rate computed from our models.

\begin{figure*}\centering
\includegraphics[width=0.35\textwidth]{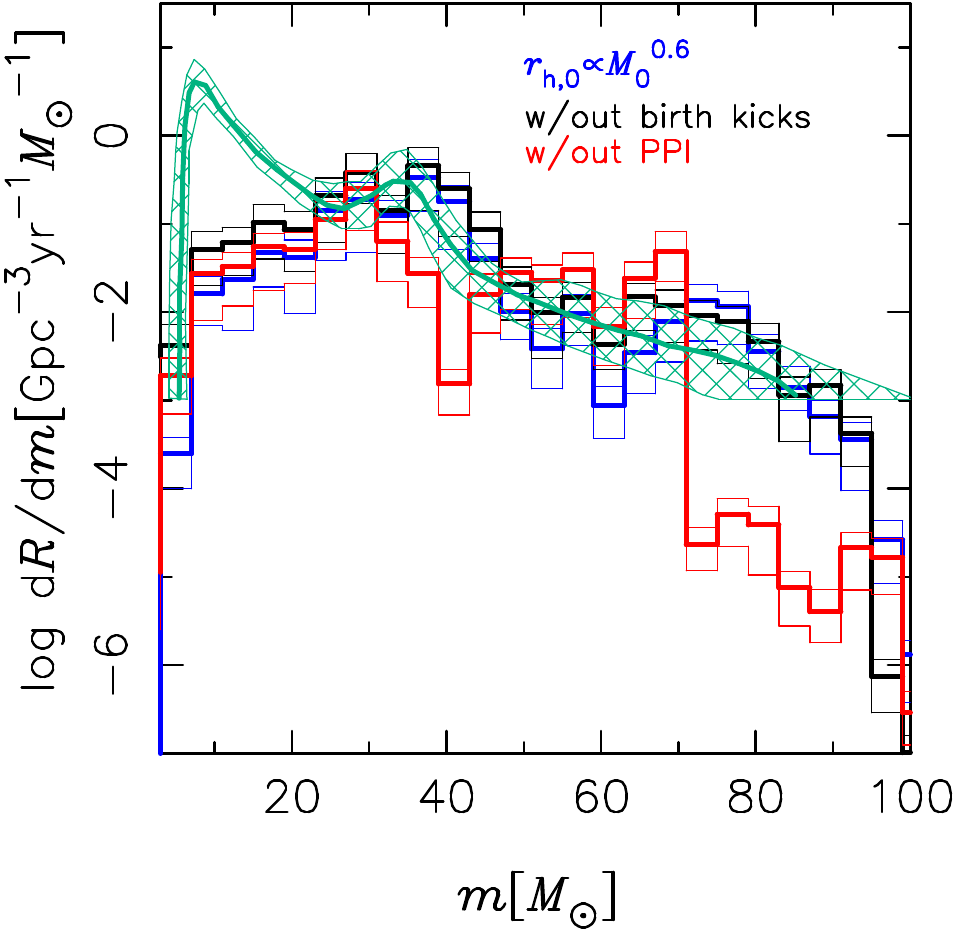}
\includegraphics[width=0.34\textwidth]{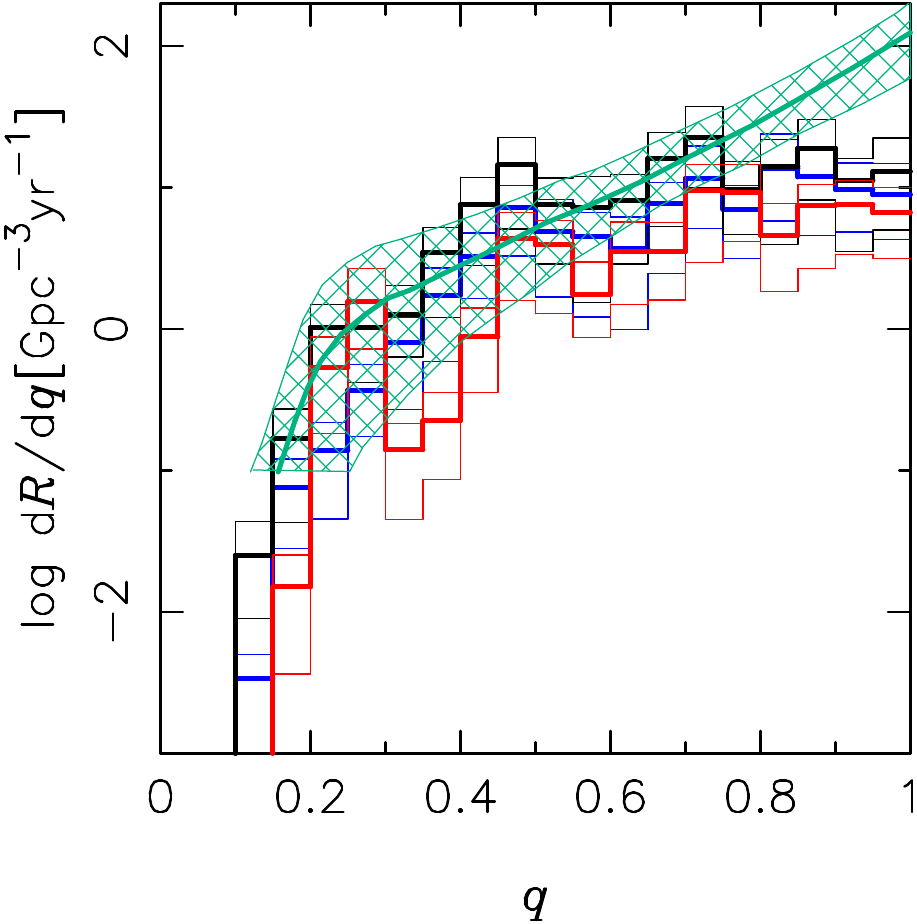}
\includegraphics[width=0.35\textwidth]{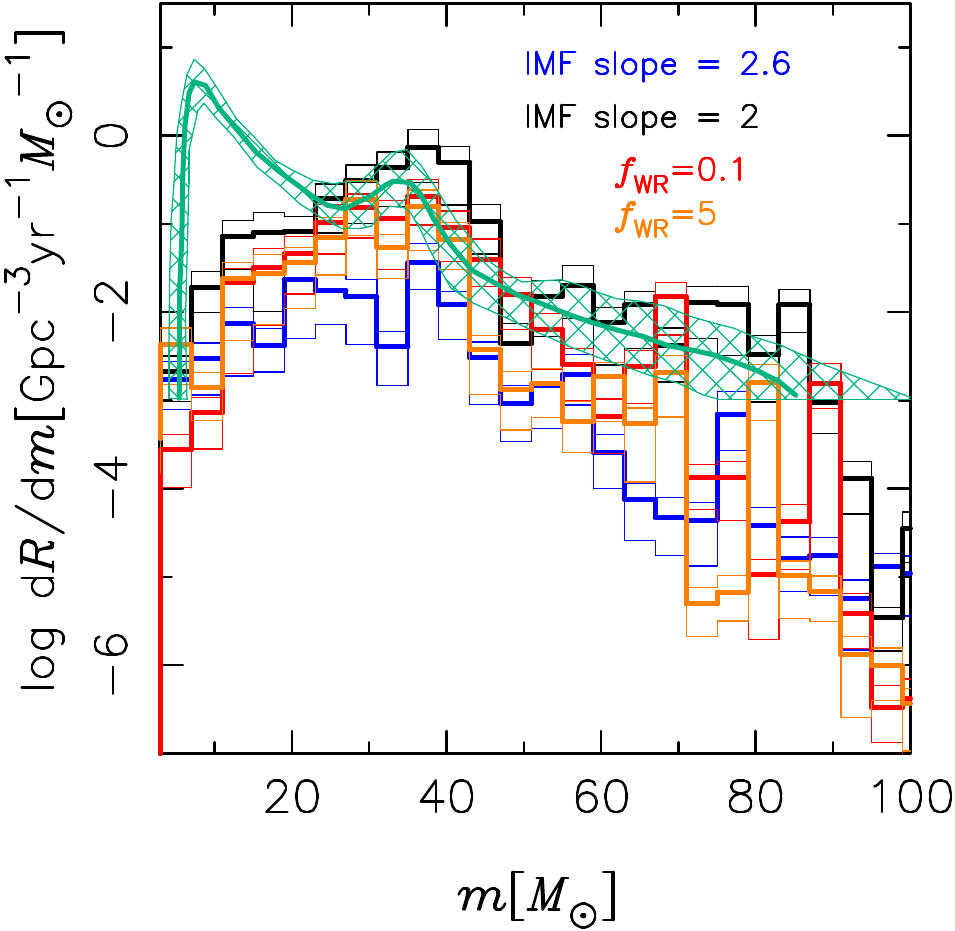}\includegraphics[width=0.34\textwidth]{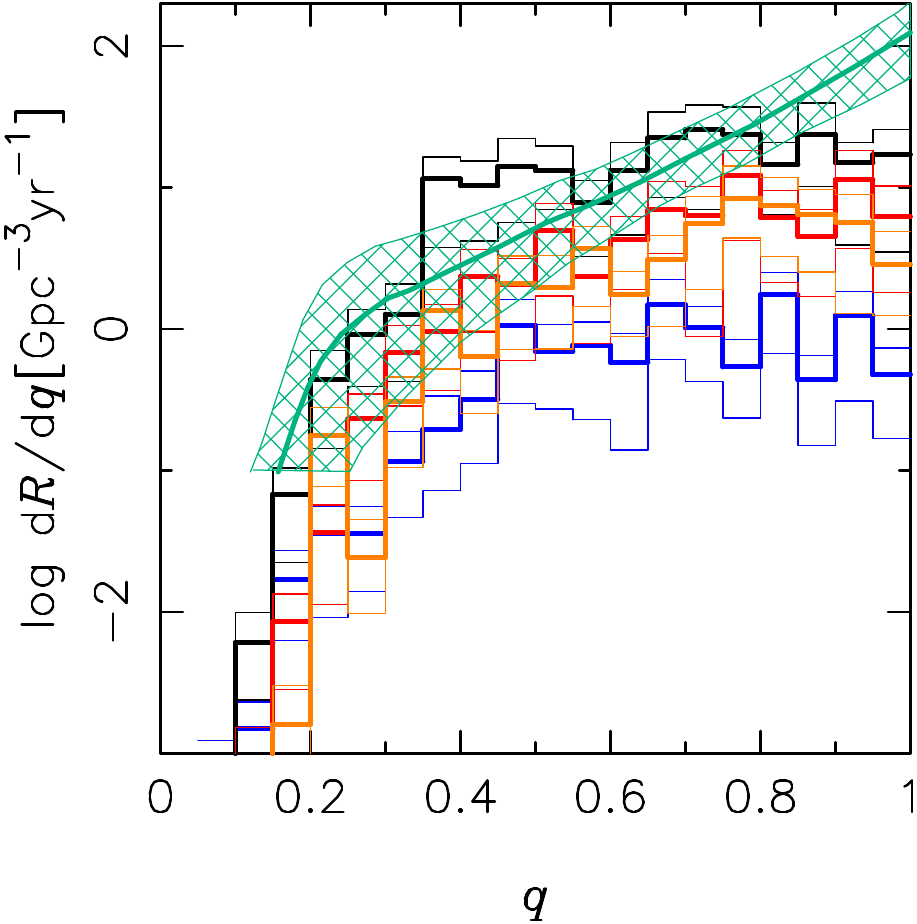}
\caption{{   Results for six alternative models. Left panels: blue histograms are for a model in which the initial cluster
half-mass radius is assumed to scale with  cluster mass as in equation~(\ref{mrr}); the blue histograms correspond to a model in which the BH birth kicks are zero; the red histograms correspond to a model  where the recipes for pulsation pair instability (PPI) were switched off. 
Right panels: blue and black histograms are the results obtained assuming that the  initial stellar mass function for massive stars scales as $\phi(m_\star)\propto m_\star^{-2}$ and  $\phi(m_\star)\propto m_\star^{-2.6}$, respectively.
In the red and orange histograms we have multiplied our standard wind mass loss rate on the Wolf-Rayet  stage by a factor $f_{\rm WR}=0.1$ and $f_{\rm WR}=5$. We have used the delayed supernova mechanism, and,
unless otherwise specified, all the other model parameters are the same as in Fig.~\ref{nospin}.}
 }\label{alter}
\end{figure*}

\subsection{Effect of initial cluster density, initial spins, and other model variations}

The number of heavier BHs produced by a cluster through hierarchical mergers is affected 
by both the cluster density and the initial spin of the BHs. 
A larger cluster density means a  larger merger rate and escape velocity, and therefore a larger probability that a remnant BH is retained inside the cluster following a recoil kick. Similarly, if the BHs have negligible spins, this translates
into a smaller recoil velocity and higher retention probability. While in the previous section we have looked at a somewhat optimistic scenario in which clusters all form with high densities and the BHs have initially zero spins, in this section we vary these assumptions and investigate their effect on the BH binary merger rate and properties.
We adopt here the delayed supernova mechanism, but similar results are obtained with the rapid supernova prescription and the \cite{Belczynski2008} mass distribution.

In the upper panels of Fig. \ref{density}, we vary the initial cluster half-mass density within the range  $\rho_{\rm h,0}=10^2$ to $10^5~\msun\, \pc^{-3}\ $, and assume that the BHs have zero spins initially.
The results illustrate that although our models can in principle account for most mergers above $m_1 \gtrsim 20M_\odot$, this is only true under some specific conditions. As we lower the initial cluster half-mass density the  merger rate goes down significantly  at all values of mass and mass-ratio. The depletion is more significant at masses above the cut-off mass of $50M_\odot$ and for $q\lesssim 0.5$.  Thus, a scenario where most merging BH binaries with $m_1 \gtrsim 20M_\odot$ form in GCs would imply a typical initial cluster density 
$\rho_0\gtrsim 10^4~\msun\, \pc^{-3}$. It is important to note that this condition would however only apply to clusters with initial mass  $M_0\gtrsim 5\times 10^4M_\odot$, where most of the merging binaries are formed (see Fig.~\ref{clusterD}).

In the lower panels of Fig. \ref{density} we show how the results change with changing the initial BH spins. In these models we keep the initial density to the fixed value $\rho_0= 10^5~\msun\, \pc^{-3}$. We see that the merger rate density distributions are not affected significantly for  $m_1\lesssim 50M_\odot$ and $q\gtrsim 0.5$. This is  because the majority  of these binaries are made of first generation BHs. Hence, their merger rate is not affected by the recoil kick velocity and  by the initial choice of BH spin.
 On the other hand,   
  the number of BHs formed via hierarchical mergers decreases significantly  when higher initial spins are used due to the larger recoil kicks. This leads to a lower merger rate at $m_1\gtrsim 50M_\odot$  when $\chi$ is increased.  Even relatively modest initial spins, $\chi=0.1$, lead to a distribution that does no longer match the inferred distribution. 
The constrains on $\chi$ seems therefore quite strong as a hierarchical  origin for all mergers with $m_1\gtrsim 50M_\odot$ would require that   BHs are formed with nearly zero spin.

{   Finally, we consider six additional model realisations. In one model, we assume that the BHs receive no kick at formation and that the initial density is the same for all clusters, $\rho_{\rm h ,0}=10^5~\msun\, \pc^{-3}\ $.
 In another model, we
assume that   the cluster half-mass radius scales  as
\begin{equation}\label{mrr}
  \log\left({{r}_{\rm h,0} \over {\rm pc} }\right)=-3.56+0.615 \log \left( {M_0\over M_\odot }\right) \ .
\end{equation}
This latter relation  {was derived by \cite{2010MNRAS.408L..16G} from the results of}  \citet{2005ApJ...627..203H} {who fit this Faber-Jackson-like relation  to  ultra-compact dwarf galaxies (UCDs) and elliptical galaxies. \citet{2010MNRAS.408L..16G} derived the initial mass-radius relation correcting for mass loss and expansion by stellar evolution and correcting radii for projection.}
We consider an additional model realisation where we did not include any prescription for pair instability  so that the initial BH mass function has no upper gap and BHs can form above $50M_\odot$.
Moreover, we consider two models where the initial mass function above $0.5\ M_\odot$ is assumed to scale  as $\phi(m_\star)\propto m_\star^{-2}$ (top-heavy) and  $\phi(m_\star)\propto m_\star^{-2.6}$ (bottom heavy), respectively.
Finally, we evolve two additional models where our standard  Wolf-Rayet winds based on  \cite{1998A&A...335.1003H} and  \cite{2005A&A...442..587V} are
multiplied by a factor $f_{\rm WR}=0.1$ and  $f_{\rm WR}=5$ \citep[e.g.,][]{2022MNRAS.tmp.1776B}.
Unless otherwise specified, all the other model parameters are the same as before, i.e., delayed supernova mechanism, $\chi=0$, fallback kicks, etc.
}

 Fig.~\ref{alter} shows  that the mass properties of the BH binaries produced in the  {new}{} models without birth kicks and with the new $r_{\rm h}$-$M$ relation are similar to those found previously in Section~\ref{fidmod}.
The fact that adopting the mass-radius relation equation~(\ref{mrr}) does not change significantly our results is not surprising. 
Clusters with an initial mass $M_0\sim 10^6M_\odot$ contribute the most to the merger rate (see Fig.\ref{clusterD}). 
The initial half-mass density of these clusters as derived from equation equation~(\ref{mrr}) is $\rho_{\rm h ,0}\simeq 5\times 10^4~\msun\, \pc^{-3}\ $. This is comparable to the constant density value of $10^5~\msun\, \pc^{-3}$ adopted previously.
Interestingly, the results of models with no birth kicks show that assuming zero velocity kicks at birth increases slightly the merger rate at the lower  mass end of the $m_1$ distribution and the number of merging binaries with asymmetric masses. On the other hand, the shape and normalisation of the distributions at masses higher than $m_1\gtrsim 20M_\odot$  remain virtually the same as in the fallback kick model.

The  model without pair instability physics leads to a mass distribution which is significantly different from the other model realisations, showing how our results can depend on the assumptions about stellar evolution and the adopted prescriptions. In this case, the mass distribution still peaks at $m_1 \sim 30~M_\odot$, while the other peak near $40~M_\odot$ is not longer present.  A secondary peak is found near $70M_\odot$.  For masses larger than this value, the merger rate drops and becomes much smaller than the  rate inferred from the GW data.

{   In  the model with a top-heavy stellar mass function, the overall merger rate  is higher than
for our standard models due to the larger number of BHs formed. On the other hand, for a bottom-heavy mass function the total merger rate is  significantly reduced due to the fewer massive stars formed.  Our model with  modified Wolf-Rayet wind mass-loss rate  lead to results that are qualitatively similar to those obtained under our more standard assumptions.

 That our resulting mass distributions are sensitive to the initial BH mass function, and therefore to the uncertain  stellar evolution prescriptions is expected. 
 It is interesting, however, that most of our models share  similar properties. Specifically: (i) the inferred peak in the merger rate at $10M_\odot$ is much lower than the one inferred from the data, and (ii) the distribution of $m_1$ presents a main peak at near $35M_\odot$.}
{ The main reason why there are so few mergers with small masses is because of the relatively low number of light BHs in the initial mass function. This is due to the low metallicity of GCs, which results in low wind mass loss and large BH masses. The other reason why the mass distribution of merging binaries peaks at relatively high values is
dynamics. The masses of the binary components tend to be sampled near the top end of the BH mass function, due to the high value of the power law exponents that appear in the density probability functions $p_1$ and $p_2$ (see Section \ref{masses}). On the other hand, the flatter $p_3$ distribution means that the ejected BH interlopers will  be on average lighter than the binary BH components. These lighter BHs are therefore no longer available for merging.  
  }

 \section{conclusions}\label{conc}
In this work we have used our fast cluster evolution code, {\tt cBHBd}, to investigate the mass distribution of BH binaries produced dynamically in dense GCs.
We compared our results to the astrophysical distribution of BH binary masses inferred from GWTC-3 to make inference about the astrophysical origin of the sources.
For the first time, we have included hierarchical mergers in our models. 
This allowed us to address the question of whether a dynamical formation scenario  is 
a feasible explanation for the detected BHs within the  so called `upper mass gap'. Such a mass gap in the initial BH mass function 
is predicted by stellar evolution theories, 
 and  in our models is located at $\gtrsim 50M_\odot$. Because {\tt cBHBd} is highly efficient compared to other techniques (e.g., Monte Carlo, $N$-body), we were able to systematically investigate the impact of model assumptions on or results.
Our main conclusions are summarised below:
\begin{itemize}
\item[i)]  A purely GC formation scenario 
for the BH binaries detected by LIGO and Virgo is inconsistent
with the $\simeq 10M_\odot$  peak in the primary BH mass distribution that is
inferred from the data. This likely excludes a scenario where the majority of the sources were formed in GCs.
\item[ii)] A GC origin can easily account for the secondary mass peak at $m_1\simeq 35M_\odot$ inferred from the  data. This requires  that  clusters form with initial half-mass density   $\gtrsim 10^4~\msun\, \pc^{-3}\ $. Assumptions about the initial BH spins and the  supernova mechanism  have no effect on this conclusion.
\item[iii)] 
 Dynamical formation in GCs can explain the inferred merger rate of  all BH binaries with  $m_1\gtrsim 20M_\odot$ and
 $q\lesssim 0.8$,  including binaries with component masses lying 
 above the assumed mass limit due to pair--instability.
For this to be true we require that both the most massive GCs, $M_0\gtrsim 10^5M_\odot$, form with half-mass density  $\gtrsim 10^4~\msun\, \pc^{-3}\ $, and that the birth spins of BHs  are nearly zero.
 Even small deviations from this latter condition lead to a merger rate above $50M_\odot$ that is orders of magnitude smaller than  the inferred rate.
 \item[iv)] A hierarchical merger scenario predicts the appearance of multiple  peaks in the primary BH mass distribution and
 within the upper mass gap 
 due  to  a pile--up of mergers between first and second generation BHs. Inter--generation mergers lead to a simple relation between the mass value of any of such peaks 
 and that of peaks found at masses lower than the  pair-instability  limit.
 These features can be tested against future GW data to place constrains on a GC origin for the sources.
\end{itemize}

 Additional constraints on the formation of BH mergers can  be placed by exploring correlations between binary parameters, which we have not considered here, but plan to study in a future work. For example, a BH formed from  a previous merger will have
a spin $\chi\simeq 0.7$. We expect therefore  a change
in the value of the typical effective and precession spin parameters of  binaries with components within the upper mass gap \citep[e.g.,][]{2021MNRAS.507.3362T,2020PhRvD.102d3002B}  and an increase in spin magnitude for systems with more unequal mass ratio.
 Binaries formed dynamically will also have larger eccentricities, which can lead to a positive correlation between eccentricity, spin and binary mass in the overall population.      
The analysis of the data from GWTC-3 has shown marginal evidence that the spin distribution broadens above $30M_\odot$, and that the mass ratio and spin are correlated in the sense that  spins are larger for more asymmetric binaries \citep{2021arXiv211103606T}. The evidence for these correlations remain weak, but it suggests  that future analysis based on larger data sets will soon be able to provide more stringent constrains. 
The residual eccentricity of a binary  is by itself another potentially powerful tool for identifying sources formed in clusters. 
\cite{2022arXiv220614695R} suggest that 
      a  significant fraction of the detected GW sources in GWTC-3 show  support for eccentricity $\gtrsim 0.1$ at $10$Hz. Their results  indicate that densely--populated star clusters may produce
the majority of the observed mergers.

Finally, it is worth noting that in our work we used the pair--instability  prescriptions  from \cite{2017MNRAS.470.4739S}.
This gives an upper limit in the initial BH mass function of about $50M_\odot$. However, there are several uncertainties in the modelling, and different assumptions can lead to significantly different values for the high mass cut-off, generally in the range $40M_\odot$ to  $70M_\odot$ \citep{2018MNRAS.474.2959G,2020ApJ...902L..36F,2021MNRAS.501.4514C,2022ApJ...931...94F}. Exploring the effect of these uncertainties is beyond the scope of this paper, but should be considered in future work.

\section*{Acknowledgements}
Some of the processing
of the results has been done using the {\sc python} programming
language and the following open source modules: {\sc
  numpy}\footnote{http://www.numpy.org}, {\sc
  scipy}\footnote{http://www.scipy.org}, {\sc
  matplotlib}\footnote{http://matplotlib.sourceforge.net}.
  We acknowledge the support of the Supercomputing Wales project, which is part-funded by the European Regional Development Fund (ERDF) via Welsh Government.  FA acknowledges support from a Rutherford fellowship (ST/P00492X/1) from the Science and Technology Facilities Council.
  MG acknowledges support from the Ministry of Science and Innovation (EUR2020-112157, PID2021-125485NB-C22, CEX2019-000918-M funded by MCIN/AEI/10.13039/501100011033) and from AGAUR (SGR-2021-01069).


\section*{Data availability}
The data underlying this article will be shared on a reasonable request to the corresponding author.

\bibliographystyle{mn2e}
\bibliography{Mendeley.bib}

\appendix

\begin{figure*}
        \includegraphics[width=0.45\textwidth]{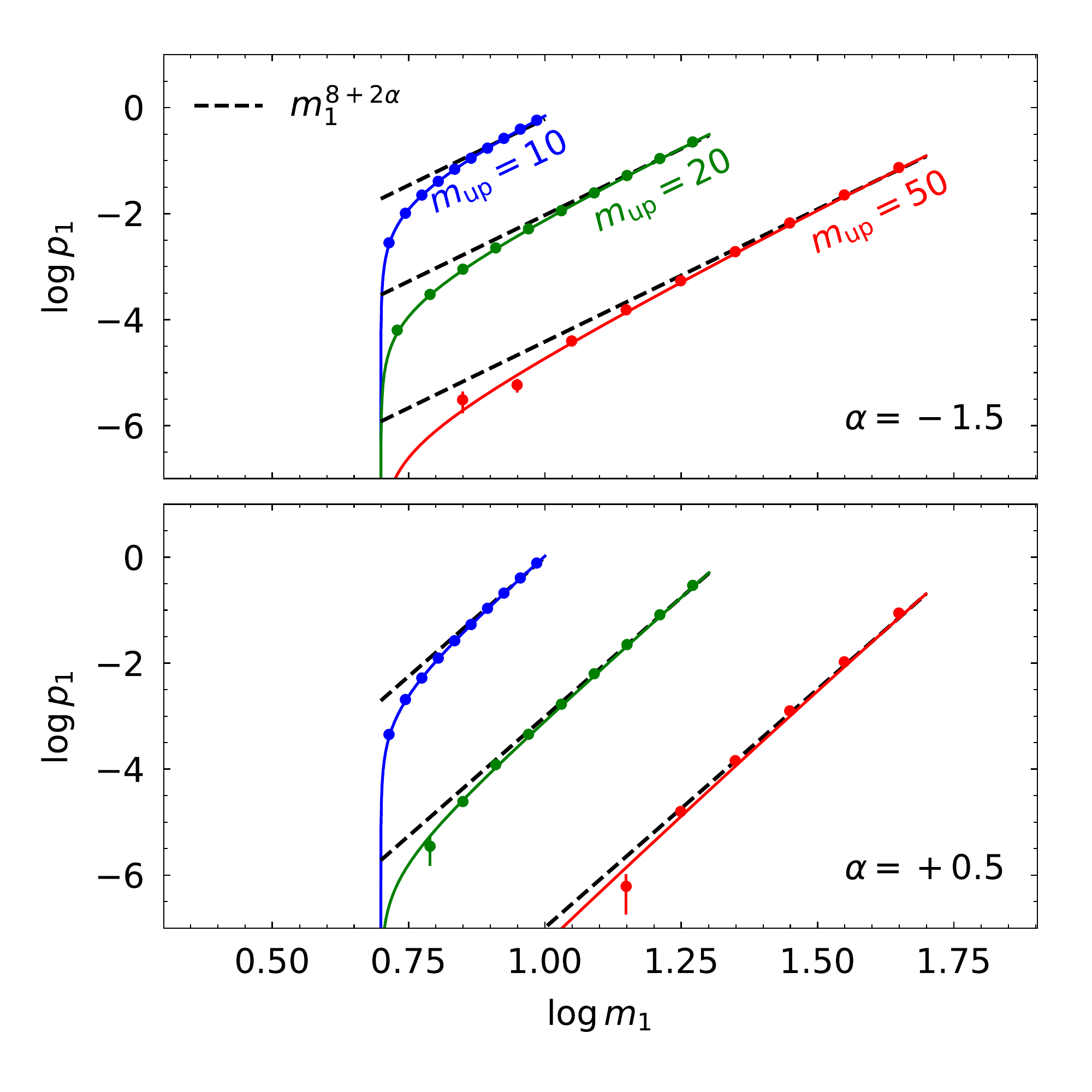}
        \includegraphics[width=0.45\textwidth]{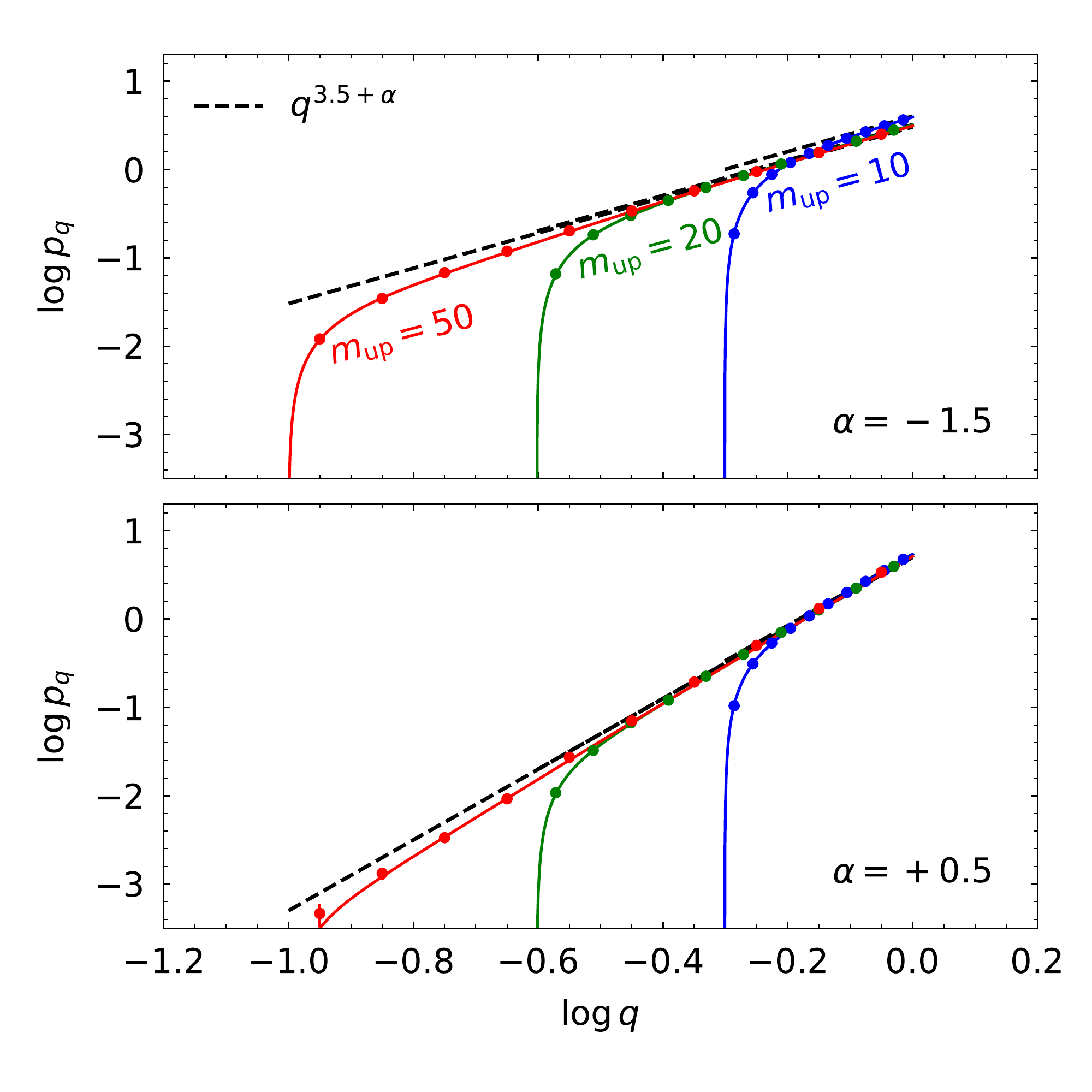}\\
        \caption{{PDFs for $m_{1}$ (left) and  $q$ (right) for different BH mass function upper masses ($m_{\rm up}$) and logarithmic slopes  ($\alpha$). In all cases $\mlo=5$. The results discussed in the text (equations~\ref{eq:pm1} and \ref{eq:pq}) are shown as full lines, while the dots with errors bars show Monte Carlo realisations (as a check) obtained by drawing $10^6$ pairs from {$p_I(m_{I})$ (equation~\ref{eq:mI})} and the black dashed lines show simple power-law approximations. }}
        \label{fig:pdfs}
\end{figure*}

\section{Mass sampling routines}\label{masses}
\subsection{Masses of 3-body binaries}
We are interested in  the probability density functions (PDFs) of the masses of the two components of (BH) binaries that form in  three-body interactions.  Following  \citet{1975MNRAS.173..729H}, the formation rate of hard binaries per unit of volume and energy is expressed as 
\begin{equation}
\Gamma_{\rm 3b}(m_I, m_{II}, m_{III}, x) = n_{I} n_{II} n_{III} Q(m_{I}, m_{II}, m_{III}, x),
\end{equation}
where $n_i = n(m_i)$ is the number density of BHs with mass $m_i$,  $m_I$ and $m_{II}$ are the masses of the stars ending up in the binary, $m_{III}$ is the mass of the catalyst star and $x$ is the (positive) binding energy of the binary. The rate function $Q$ is given by equation 4.14 in \citet{1975MNRAS.173..729H} and it is a function of the three masses, as well as their $\beta_i=(m_i\sigma_i^2)^{-1}$, where $\sigma_i$ is the one-dimensional velocity dispersion of mass component $i$. Because  binaries tend to form from the most massive objects, for which  energy equipartition is established quickly, we assume $\beta_I=\beta_{II}=\beta_{III}=\beta$. Integrating equation~4.14 from Heggie over all $x$, from the hard-soft boundary (i.e. $x=\beta^{-1}$) to $\infty$ (i.e. all hard binaries), we find that the formation rate of hard binaries per unit of volume is
\begin{equation}
\Gamma_{\rm 3b}(m_{I}, m_{II}, m_{III}) \propto n_{I} n_{II} n_{III}\frac{(m_{I}m_{II})^4m_{III}^{5/2}}{\sqrt{(m_I+m_{II}+m_{III})(m_I+m_{II})}}\beta^{9/2}.
\label{eq:gamma3b}
\end{equation}
 For equal masses, this result reduces to the frequently used scaling $\Gamma_{\rm 3b}(m)\propto n^3m^5\sigma^{-9}$. Equation~(\ref{eq:gamma3b}) is symmetric in $m_I$ and $m_{II}$, so the PDF for the mass of one of them is found from

\begin{equation}
p_I(m_{I}) =  \int_{\mlo}^{\mup} \int_{\mlo}^{\mup} \dr m_{III} \dr m_{II}\, \Gamma_{\rm 3b},
\label{eq:mI}
\end{equation}
and  $p_{II}(m_{II})=p_I(m_I)$. Here $\mlo$ and $\mup$ are the lower and upper bound of the mass distribution, respectively. 

{ We note here that
 Equation \ref{eq:gamma3b} includes the assumption of equipartition and therefore takes into account the dependence of the velocity dispersion on mass.
 On the other hand, we do not consider the change in  the BH mass function that is expected in the core due to mass segregation. Numerical simulations have shown that the mass function in the core has a logarithmic slope that is approximately only $+1$ steeper than the  global mass function 
 \citep[e.g.,][]{2007MNRAS.378L..29P}. Thus we expect the effect of mass segregation on our results to be relatively small.
 }

We now adopt the convention that $m_1$ and $m_2$ are the most massive and least massive component, respectively, with corresponding PDFs $p_1(m_1)$ and $p_2(m_2)$. These correspond to the PDFs of the maximum and minimum value, respectively, when a sample of two values are drawn from $p_I(m_{I})$, and are given by
\begin{align}
p_1(m_1) &= 2P_I(m_{1})p_I(m_{1}),\label{eq:pm1}\\
p_2(m_2) &= 2\left[1-P_I(m_2)\right]p_I(m_2),
\end{align}
where $P_I(m_{I})=\int_{\mlo}^{m_{I}} \dr m_{I}^\prime \,p_I(m_{I}^\prime)$ is the cumulative density function (CDF) of $p_I(m_{I})$. 

The PDF of $q$ is a ratio distribution and can be found from the joint distribution of the minimum and maximum values, which is given by $p_{12}(m_1, m_2) = 2p_1(m_1)p_2(m_2)$ and 

\begin{align}
p_q(q) = \int_{\mlo}^{\mup}\dr m_2 \,  p_{12}(qm_2,m_2).
\label{eq:pq}
\end{align}

We then assume that the mass function is a power law such that $n_i \propto m_i^\alpha$ between $\mlo$ and $\mup$. 
A value of $\alpha=0.5$ provides a good approximation of the mass function of BHs at low metallicities \citep[$Z\lesssim 0.05Z_\odot$, see Fig. 4 of][]{2020MNRAS.492.2936A}.
In Fig.~\ref{fig:pdfs} we show the resulting
$p_1(m_1)$ and $p_q(q)$ for $\mup=[10,20,50]$ and $\mlo=5$ and $\alpha=+0.5$ {(approximate for metal-poor GCs) and $\alpha=-1.5$ (approximate for metal-rich GCs)}. We find that these PDFs can be reasonably well approximated by power-laws of the form: $p_1(m_1)\propto m_1^{8+2\alpha}$ and $p_q(q)\propto q^{3.5+\alpha}$, for all values of $\mup$ and $\alpha$. This approximation can be used to sample $m_1$ and $m_2$ (via $q$).  

\subsection{Masses of interlopers}
Assume a binary BH with mass $M_{12}=m_1+m_2$, moving in a field of BHs with  number density $n_3$.
The  rate of encounters between the BH binaries and field BHs is \citep{1976ApL....17...87H}
\begin{equation}
\dot{N}_3 = n_3\langle \Sigma v\rangle,
\end{equation}
where $v$ is the relative velocity between the binary BH and another BH and $\Sigma$ is the cross section for an encounter, which in the gravitational focusing regime is \citep{1976ApL....17...87H}
\begin{align}
\Sigma       \simeq \frac{2\pi G a M_{123}}{v^2},
\end{align}
where $G$ is the gravitational constant{, $M_{123} = M_{12}+m_3$} and $a$ is the  semi-major axis of the binary.
We can find $\langle \Sigma v\rangle$ from integrating over all velocities

\begin{align}
\langle \Sigma v\rangle 
 &= \frac{4l^3}{\pi^{1/2}}\int_0^\infty \Sigma(v) v^3 \exp(-l^2v^2)\dr v,\\
&=4\pi^{1/2}GlM_{123}a .
\end{align}
Here $l^2 = \beta M_{12}m_3/(2M_{123})$  for our assumption of equipartition.
The semi-major axis is $a\propto Gm_1m_2\beta$ such that the interaction rate scales with the masses as
\begin{equation}
\dot{N}_3 \propto m_3^\alpha \sqrt{m_3} \sqrt{M_{12}M_{123}}m_1m_2
\end{equation}
So, interactions with more massive BHs are slightly favoured wrt random draws from the BH mass function. 
Because $p_1(m_1)$ and $p_2(m_2)$ are much steeper than this distribution, we find that to good approximation  $p_3(m_3)\propto m_3^{\alpha+1/2}$. It also means that exchanges are not very important, because these happen when the intruder is more massive than any of the binary members. {Here we find for $\mup/\mlo = 10$ and $\alpha=+0.5$ that $\langle m_1\rangle \simeq 0.91\mup; \langle m_2\rangle \simeq0.76\mup$ and $\langle m_3\rangle \simeq 0.55\mup$ and exchange interactions should therefore not be very common. Once the width of the BH mass function has shrunk to $\mup/\mlo\simeq2$, $\langle m_2\rangle\simeq0.78$ which is comparable to $\langle m_3\rangle\simeq0.75$ and exchange interactions (which we neglect) are more relevant.}


\bsp	
\label{lastpage}
\end{document}